\documentclass[twocolumn]{aastex63}
\pdfoutput=1

\newcommand\tess{{\it TESS}}
\newcommand\teff{{\log T_{\rm {eff}}}}
\newcommand\lum{{\log L/L_\odot}}
\newcommand\vsini{{v_{\rm eq}\sin i}}
\newcommand\vturb{{\zeta_{RT}}}
\newcommand\vmicro{{\xi_{\rm eff}}}
\newcommand\vfirst{{\langle v\rangle}}
\newcommand\vsecond{{\langle v^2\rangle}}
\newcommand\vthird{{\langle v^3\rangle}}

\usepackage{amsmath,graphicx,multirow,mathtools}

\defcitealias{dornwallenstein22}{DW22}
\defcitealias{chen24}{CDW24}
 
\received{23 July, 2025}
\revised{8 August, 2025}
\accepted{11 August, 2025}
\submitjournal{ApJ}

\shorttitle{Turbulence in LMC YSGs}
\shortauthors{Dorn-Wallenstein et al.}

\begin{document}

\title{A Spectroscopic Hunt for Post-Red Supergiants in the Large Magellanic Cloud II: Turbulent Line Broadening in the Spectra of LMC Yellow Supergiants}

\correspondingauthor{Trevor Z. Dorn-Wallenstein}
\email{tdorn-wallenstein@carnegiescience.edu}

\author[0000-0003-3601-3180]{Trevor Z. Dorn-Wallenstein}\thanks{Carnegie Fellow}
\affiliation{The Observatories of the Carnegie Institution for Science, 813 Santa Barbara Street, Pasadena, CA 91101, USA}

\author[0009-0007-4983-9850]{Kaitlyn M. Chen}
\affiliation{Harvey Mudd College,
340 E Foothill Blvd.,
Claremont, CA 91711, USA}
\affiliation{The Observatories of the Carnegie Institution for Science, 813 Santa Barbara Street, Pasadena, CA 91101, USA}

\author[0000-0003-2872-5153]{Samantha C. Wu}\thanks{Carnegie Fellow}
\affiliation{The Observatories of the Carnegie Institution for Science, 813 Santa Barbara Street, Pasadena, CA 91101, USA}

\author[0000-0003-1012-3031]{Jared A.~Goldberg}
\affil{Center for Computational Astrophysics, Flatiron Institute, 162 5th Ave, New York, NY 10010, USA} 

\author[0000-0002-7296-6547]{Anna J. G. O'Grady}
\affiliation{McWilliams Center for Cosmology and Astrophysics, Department of Physics, Carnegie Mellon University, Pittsburgh, PA 15213, USA}

\author[0000-0002-6838-0252]{Ayanna T. Mann}
\affiliation{Howard University, Washington, DC 20059, USA} 
\affil{Center for Computational Astrophysics, Flatiron Institute, 162 5th Ave, New York, NY 10010, USA}
\affil{Department of Astronomy and Columbia Astrophysics Laboratory, Columbia University, New York, NY 10027, USA}

\author{Poderosa I. Don-Wallanchez}
\affil{3026 E Orange Grove Blvd,
Pasadena, CA 91107, USA}

\begin{abstract}

Massive stars in the Hertzsprung gap are a mixed population of objects in short-lived evolutionary phases: yellow supergiants (YSGs) evolving towards the red supergiant (RSG) phase, partially-stripped post-RSGs, and other, rarer outcomes of stellar evolution. Studies of sufficiently large samples of these objects can constrain massive star structure and evolution during these poorly-understood phases. As part of our ongoing program searching for post-RSGs, we characterized the spectral line profiles of 32 YSGs in the Large Magellanic Cloud using high-resolution spectra obtained with the MIKE spectrograph on the Magellan 2/Clay telescope at Las Campanas Observatory. We find that the line profiles are strongly broadened by turbulent photospheric motion. After fitting the profiles to measure microturbulent and macroturbulent velocities, we identify two groups within our sample that are separated by the ratio of the two velocity scales. In both groups, the macroturbulent velocity $\vturb$ scales with stellar properties such as effective temperature. Additionally, we find statistically-significant correlations between the macroturbulent velocity and other possible probes of large-scale photospheric motions: line profile asymmetry, as well as the amplitude and quality factor of the stochastic low frequency variability measured from {\it Transiting Exoplanet Survey Satellite} lightcurves. These correlations differ between the two groups of YSGs. Finally, we construct 1D evolutionary models of YSGs in both pre- and post-RSG phases, and find reasonable agreement between the convective velocities in these models and our measured microturbulent velocities. However, the macroturbulent velocities are much higher than the convective velocities in the models. 

\end{abstract}

\section{Introduction}\label{sec:intro}

In papers, proposals, and talks, yellow supergiants (YSGs) are often referred to as ``magnifying glasses'' \citep[a phrase first coined by][]{kippenhahn90}; attempts to reconcile models with observations tend to reveal just how little is understood about many of the physical processes driving stellar evolution. While overused, this metaphor is an apt one. Massive stars first pass through the YSG phase after leaving the main sequence. The core undergoes significant structural adjustments, the stellar radius grows by orders of magnitude in response, and the envelope begins to develop large scale convection, all during an incredibly brief period of $<1$ Myr \citep{ekstrom12} corresponding to the thermal timescale of the contracting Helium core. 

Amidst these ``normal'' YSGs are interspersed massive stars with drastically different evolutionary histories. Observations have confirmed that at least some stars with masses above $\sim$20 M$_\odot$ experience a second YSG phase \citep{gordon19}, which appears to be the result of strong mass loss from either winds or binary interactions during the red supergiant (RSG) phase \citep{eldridge17,massey21,yang23,vink23}. These partially-stripped YSGs/post-RSGs appear to be the progenitors of Type IIb supernovae \citep[e.g.][]{dessart11,yoon10,yoon17,gilkis22}. Additionally, some stars whose envelopes are entirely stripped experience a ``Helium main sequence'' phase, from which they can expand and enter this region of the HR diagram as Helium giants \citep{laplace20}.

Alternatively, some YSGs may be the products of mass {\it gain} from a binary companion, or even the result of a stellar merger \citep[e.g.][]{eldridge17,schneider24,menon2024}. Or, depending on the precise treatment of convection in the deep interior of a star, some stellar evolution models predict the existence of a ``blue loop'' during the RSG phase, which may briefly produce YSG-like objects \citep{stothers91,walmswell15,wagle19}. Finally, the most luminous YSGs with $\lum\gtrsim5.4$ and $\teff\approx3.85$ can encounter the ``yellow void'', a region in the HR diagram where their envelopes become dynamically unstable, resulting in outbursts and eruptions \citep{maeder92,nieuwenhuijzen95,stothers01}. All of these evolutionary scenarios create stars with a similar range of effective temperatures and luminosities. However, the masses and structures of the envelopes of these stars may differ drastically depending on which ``type'' of YSG it is. For example, a partially-stripped YSG will have a more tenuous and lower-mass envelope than a mass-gainer YSG in the same position in the Hertzprung-Russell (HR) diagram.

These differences are encoded within the turbulent velocity fields in the outer layers, since the density, temperature, luminosity, and opacity within a given layer and its surroundings will affect the existence, strength, and morphology of convection \citep[e.g.][]{bohmvitense58,stein98,stein89}, which thus may be observable via high-resolution spectroscopy. To observationally distinguish between normal YSGs and YSGs with alternate evolutionary histories (with a particular focus on identifying post-RSGs), we began a spectroscopic survey of the most luminous YSGs in the Large Magellanic Cloud (LMC) using the Magellan Inamori Kyocera Echelle (MIKE) spectrograph. Our sample is focused on LMC cool supergiants with luminosities above $\lum=5$ \citep{neugent12_ysg}. Notably, this boundary defines the onset in luminosity space of the {\it red supergiant problem}, the statistical discrepancy between the predicted and observed progenitor masses of type II-P supernovae (\citealt{smartt15,rodriguez22,dornwallenstein23}, but see also \citealt{beasor24,healy24}), which may be resolved by invoking strong mass loss in luminous RSGs. 

The nominal goal of our survey was to examine the properties of known LMC post-RSGs \citep{humphreys23}, determine if they were distinguishable in our data, and identify new post-RSGs to determine the frequency with which massive stars undergo such blueward evolution prior to core-collapse. In Paper I of this series, \citet{chen24} (\citetalias{chen24} hereafter) presented a preliminary analysis of our MIKE data, including measurements of $\log g$ and $T_{\rm eff}$ using a template-fitting approach. They found that many of the resulting surface gravity measurements were anomalously high for cool supergiants, which generally should have $\log g \lesssim 2$. However, many of the spectra were best fit by models with much higher surface gravities, such that some of the masses inferred by combining the measured atmospheric parameters were well in-excess of 100 $M_\odot$, a highly unlikely result. However, this finding can be neatly explained if the spectra of the stars in our sample were broadened by a physical process not included in the ATLAS9 template spectra used by \citetalias{chen24}, namely turbulent broadening. 

Macroturbulence --- i.e. turbulent motion on size scales greater than the photon mean free path --- is ubiquitous in the massive star regime, including cool supergiants \citep[c.f. Fig. 17.10 of][]{gray05}. Thanks to large spectroscopic surveys of OB stars like the IACOB project \citep{simon-diaz14}, studies such as \citet{simon-diaz17} and \citet{bowman20} have identified statistically significant correlations between macroturbulent velocity and effective temperature, luminosity, line asymmetry, and the stochastic low frequency (SLF) variability detectable in the short-cadence lightcurves of all massive stars \citep{blomme11,bowman19b,dornwallenstein20b,pedersen25}. However, the physical mechanism responsible for the large-scale mostly-horizontal turbulent motions invoked by the macroturbulence model is not currently known. Most recently, \citet{serebriakova24} studied a large sample of OB stars in the Galaxy and the LMC, and proposed that macroturbulence in stars with initial masses above 30 $M_\odot$ is driven by subsurface convection, consistent with theoretical predictions by \citet{cantiello21,schultz22}, and others. However, in less massive stars, it may be linked to internal gravity waves excited at the interface between the convective core and the radiative envelope \citep[see, e.g., discussion in][]{bowman19b}, up to a certain amplitude \citep[e.g.][]{anders23}. Of course, the structures of OB stars are drastically different from cool supergiants, and macroturbulence in YSGs is likely driven by a different mixture of these or other physical processes --- particularly in the context of SLF variability, convection in the envelopes of YSGs would damp internal gravity waves.

Our survey of YSGs can therefore examine how turbulence evolves as stars cross the Hertzprung gap, painting a comprehensive picture of turbulent velocity fields in the upper HR diagram, which can be compared to past results in the OB star regime. The paper is laid out as follows: In \S\ref{sec:sample}, we describe our sample in more detail, including the first full description of our MIKE observations and an overview of our previous estimates of the atmospheric parameters of our targets. \S\ref{sec:ACID} details our methodology for measuring average spectral line profiles for each star using the publicly available Python package {\sc ACID} \citep{dolan24}, our approach to measuring the statistical moments of these profiles to probe assymetries, and how we measure macroturbulent and microturbulent velocities. In \S\ref{sec:results}, we analyze our findings, comparing our measurements with the properties of the stars in our sample, the shapes of the derived line profiles, as well as measurements of the stochastic low frequency (SLF) variability in the {\it Transiting Exoplanet Survey Satellite} (\tess~) lightcurves of our targets. We discuss our findings within the broader context of stellar evolution, and present a comparison with 1D stellar evolution models in \S\ref{sec:discuss} before summarizing and concluding in \S\ref{sec:conclusion}.

\section{Sample and Observations}\label{sec:sample}

Our sample of stars is derived from \citet{dornwallenstein22} (\citetalias{dornwallenstein22} hereafter), who analyzed the \tess~lightcurves of 126 cool supergiants in the Magellanic Clouds. We chose to focus exclusively on the LMC, as these stars are in the \tess~southern continuous viewing zone, and thus have significantly more coverage. We selected all LMC stars from \citetalias{dornwallenstein22}; observations were done in order of descending photometric luminosities (derived by \citealt{neugent12_ysg} using 2MASS photometry). All stars in the \citetalias{dornwallenstein22} sample above $\lum=5$ were observed. Table \ref{tab:sample} shows the common names, coordinates, and $V$-band magnitudes of the 40 stars in this initial sample. 

\begin{deluxetable*}{llll}
\tabletypesize{\scriptsize}
\tablecaption{Names, coordinates, and $V$ magnitudes of the stars in our sample, sorted by $V$ magnitude.\label{tab:sample}}
\tablehead{\colhead{Common Name} & \colhead{R.A.} & \colhead{Dec} & \colhead{$V$}\\
\colhead{} & \colhead{[hours]} & \colhead{[degrees]} & \colhead{[mag]}} 
\startdata
HD 33579 & 05:05:55.5114 & -67:53:10.932 & 9.13 \\ 
HD 269781 & 05:34:22.4676 & -67:01:23.569 & 9.87 \\ 
HD 269723 & 05:32:24.9642 & -67:41:53.614 & 9.91 \\ 
HD 269953 & 05:40:12.1666 & -69:40:04.853 & 9.95 \\ 
HD 268819 & 04:55:32.4586 & -69:57:45.085 & 10.08 \\ 
HD 269902 & 05:38:09.5829 & -69:06:21.316 & 10.23 \\ 
HD 269697 & 05:31:38.4223 & -67:28:11.562 & 10.28 \\ 
HD 269662 & 05:30:51.4749 & -69:02:58.590 & 10.28 \\ 
HD 269857 & 05:36:32.3788 & -68:54:01.671 & 10.29 \\ 
HD 269331 & 05:18:01.8324 & -69:33:37.767 & 10.31 \\ 
HD 269840 & 05:36:10.0810 & -68:55:41.264 & 10.33 \\ 
HD 269661 & 05:30:50.0811 & -69:31:29.399 & 10.34 \\ 
HD 268946 & 05:05:12.2203 & -66:44:12.572 & 10.42 \\ 
CD-69 310 & 05:30:02.2715 & -67:02:45.077 & 10.70 \\ 
HD 269787 & 05:34:34.3613 & -66:58:23.423 & 10.73 \\ 
HD 268687 & 04:50:55.8562 & -69:25:52.504 & 10.73 \\ 
HD 269651 & 05:30:32.4335 & -69:09:11.891 & 10.73 \\ 
HD 269604 & 05:28:31.3672 & -68:53:55.749 & 10.74 \\ 
HD 269982 & 05:40:57.8100 & -69:15:31.143 & 10.79 \\ 
HD 269110 & 05:09:10.6083 & -69:36:12.203 & 10.82 \\ 
HD 270050 & 05:42:29.1686 & -67:19:41.961 & 10.84 \\ 
HD 269807 & 05:35:00.2460 & -67:01:12.688 & 10.85 \\ 
SP77 31-16 & 04:54:36.8435 & -69:20:22.179 & 10.89 \\ 
SK -69 148 & 05:28:27.9201 & -69:12:57.413 & 10.93 \\ 
HD 269879 & 05:36:47.1963 & -66:45:46.005 & 10.94 \\ 
HD 269070 & 05:05:58.1248 & -70:32:05.621 & 11.06 \\ 
{[W60]} D17 & 05:30:02.2715 & -67:02:45.077 & 11.29 \\ 
CPD-69 496 & 05:40:59.1411 & -69:20:35.438 & 11.30 \\ 
HD 269762 & 05:34:09.9708 & -68:59:12.649 & 11.37 \\ 
HD 268865 & 04:59:43.5492 & -68:31:22.779 & 11.46 \\ 
RM 1-77 & 04:56:49.4338 & -69:48:31.512 & 11.55 \\ 
HD 268727 & 04:56:39.9075 & -66:44:36.703 & 11.59 \\ 
HD 268971 & 05:01:45.1929 & -70:35:52.163 & 11.61 \\ 
 HV 883 & 05:00:07.5644 & -68:27:00.064 & 11.63 \\ 
HD 268949 & 05:03:13.6191 & -68:33:35.382 & 11.76 \\ 
HD 268828 & 04:56:40.0088 & -69:41:56.356 & 11.80 \\ 
2MASS J05344326-6704104 & 05:34:43.2717 & -67:04:10.462 & 11.92 \\ 
SK -69 99 & 05:18:30.1694 & -69:13:14.054 & 11.94 \\ 
SP77 48-6 & 05:21:46.9782 & -71:19:43.909 & 12.00 \\ 
 HV 2450 & 05:19:53.2668 & -68:04:03.693 & 12.00 \\ 
\enddata
\end{deluxetable*}

\subsection{MIKE Observations}\label{subsec:MIKE}

While our observations were discussed in \citetalias{chen24}, the limitations of the research note format did not allow for a full description of the properties of the instrument and our observational strategy, which we include here for completeness. We proposed for and obtained MIKE observations of the stars in our sample during the 2022B semester. MIKE is a cross-dispersed double echelle spectrograph mounted on the Clay/Magellan2 telescope at Las Campanas Observatory that is fully described in \citet{bernstein03}. With blue and red arms, MIKE simultaneously covers the wavelength range from roughly 3350-5000/4900-9500\AA (blue/red respectively.) MIKE offers a variety of spectral resolutions depending on the chosen setup. With the 0.7x5'' slit and 2 (spatial) by 2 (spectral) binning, we obtained $R\sim45500/58000$, corresponding to a velocity resolution of 6.6/5.2 km s$^{-1}$ (blue/red respectively).

Observations were carried out on the nights of 26 and 27 December, 2022 and 3 January, 2023. During daytime calibrations, we obtained bias frames (unused by the reduction pipeline discussed below), an exposure of the internal Th-Ar lamp for wavelength calibration, exposures of the internal quartz lamp for mapping of the spectral orders, and exposures of the internal quartz lamp with a diffuser slide inserted for flat-fielding (``milky flats''). Because the reddest MIKE orders are subject to fringing, we also obtained dome flats, which were illuminated by external quartz lamps and a variable lamp controlled by a potentiometer in the dome. All exposures were read out in the slow mode. 

At night, targets brighter than $V=11$ were observed with 3x300s exposures. Fainter stars were observed with 3x600s exposures. While seeing was generally between 1-1.2'', during a brief window on 26 December, seeing rose above 1.5''. As a result, we observed two targets (HD 269661 and HD 269662) with 3x450s exposures. We took additional Th-Ar lamp exposures every hour through the night to accommodate for time-dependent shifts in the wavelength solution of the spectrograph.\footnote{The instrument is decoupled from the telescope itself, making the wavelength solution dependent only on conditions in the dome, and not the orientation of the telescope.} Finally, at various points through the night we observed the flux standards Feige 101, GD 71, and HR 3454. Data were reduced using {\sc CarPy} \citep{kelson00,kelson03}, a tailor-made python package for the reduction of data from Magellan instruments. The MIKE pipeline ultimately produces sky-subtracted, wavelength-calibrated spectra for each spectral order corrected for flat fielding, fringing, and the blaze function of the spectrograph. 

\begin{deluxetable*}{lllllllr}
\tabletypesize{\scriptsize}
\tablecaption{Updated atmospheric parameters for the stars in our sample, as well as derived quantities, sorted by $T_{\rm eff}$. Stars with $T_{\rm eff} > 10^4$ (which are excluded from further analysis) are indicated.\label{tab:atmoparam}}
\tablehead{\colhead{Common Name} & \colhead{$T_{\rm eff}$} & \colhead{$\log g$} & \colhead{$R$} & \colhead{$\lum$} & \colhead{$\log \mathcal{L}/\mathcal{L}_\odot$} & \colhead{$E(B-V)$} & \colhead{Excluded?}\\
\colhead{} & \colhead{[K]} & \colhead{[cgs]} & \colhead{[$R_\odot$]} & \colhead{} & \colhead{} & \colhead{[mag]} & \colhead{}} 
\startdata
 HV 883 & $4377^{+80}_{-70}$ & $0.27^{+0.26}_{-0.14}$ & $282^{+8}_{-6}$ & $4.419^{+0.029}_{-0.027}$ & $3.70^{+0.15}_{-0.27}$ & $0.06^{+0.03}_{-0.03}$ &  \\ 
{[W60]} D17 & $4439^{+52}_{-57}$ & $0.06^{+0.07}_{-0.04}$ & $648^{+12}_{-11}$ & $5.166^{+0.021}_{-0.024}$ & $3.92^{+0.05}_{-0.08}$ & $0.28^{+0.02}_{-0.03}$ &  \\ 
2MASS J05344326-6704104 & $4472^{+36}_{-50}$ & $0.09^{+0.09}_{-0.06}$ & $545^{+9}_{-8}$ & $5.028^{+0.017}_{-0.018}$ & $3.90^{+0.07}_{-0.10}$ & $0.28^{+0.02}_{-0.02}$ &  \\ 
SP77 31-16 & $4491^{+33}_{-48}$ & $0.08^{+0.09}_{-0.06}$ & $922^{+15}_{-14}$ & $5.492^{+0.018}_{-0.018}$ & $3.92^{+0.06}_{-0.10}$ & $0.35^{+0.02}_{-0.02}$ &  \\ 
SK -69 148 & $4494^{+19}_{-23}$ & $0.03^{+0.05}_{-0.02}$ & $474^{+6}_{-6}$ & $4.914^{+0.013}_{-0.008}$ & $3.97^{+0.03}_{-0.05}$ & $0.01^{+0.01}_{-0.01}$ &  \\ 
RM 1-77 & $4549^{+27}_{-27}$ & $0.23^{+0.15}_{-0.12}$ & $415^{+8}_{-6}$ & $4.823^{+0.016}_{-0.016}$ & $3.79^{+0.12}_{-0.16}$ & $0.22^{+0.02}_{-0.02}$ &  \\ 
SP77 48-6 & $4941^{+57}_{-57}$ & $2.23^{+0.32}_{-0.27}$ & $672^{+10}_{-10}$ & $5.384^{+0.017}_{-0.017}$ & $1.93^{+0.27}_{-0.31}$ & $0.63^{+0.02}_{-0.02}$ &  \\ 
HD 269723 & $4958^{+19}_{-18}$ & $0.02^{+0.03}_{-0.01}$ & $741^{+4}_{-4}$ & $5.476^{+0.009}_{-0.008}$ & $4.15^{+0.02}_{-0.03}$ & $0.07^{+0.01}_{-0.01}$ &  \\ 
 HV 2450 & $4976^{+140}_{-134}$ & $2.80^{+0.14}_{-0.26}$ & $857^{+36}_{-32}$ & $5.609^{+0.019}_{-0.018}$ & $1.38^{+0.27}_{-0.15}$ & $1.15^{+0.04}_{-0.03}$ &  \\ 
HD 269879 & $5005^{+8}_{-10}$ & $0.01^{+0.02}_{-0.01}$ & $399^{+2}_{-2}$ & $4.953^{+0.003}_{-0.003}$ & $4.17^{+0.01}_{-0.03}$ & $0.00^{+0.00}_{-0.00}$ &  \\ 
HD 269110 & $5185^{+18}_{-16}$ & $0.80^{+0.11}_{-0.11}$ & $362^{+3}_{-3}$ & $4.929^{+0.007}_{-0.004}$ & $3.45^{+0.11}_{-0.11}$ & $0.00^{+0.01}_{-0.00}$ &  \\ 
HD 269070 & $5191^{+17}_{-15}$ & $0.71^{+0.10}_{-0.10}$ & $353^{+3}_{-3}$ & $4.910^{+0.006}_{-0.004}$ & $3.54^{+0.10}_{-0.10}$ & $0.00^{+0.01}_{-0.00}$ &  \\ 
HD 268828 & $5340^{+23}_{-20}$ & $1.22^{+0.13}_{-0.12}$ & $210^{+2}_{-2}$ & $4.510^{+0.014}_{-0.013}$ & $3.09^{+0.12}_{-0.13}$ & $0.02^{+0.01}_{-0.01}$ &  \\ 
HD 268865 & $5518^{+13}_{-12}$ & $1.44^{+0.12}_{-0.12}$ & $216^{+1}_{-1}$ & $4.590^{+0.006}_{-0.003}$ & $2.92^{+0.12}_{-0.12}$ & $0.00^{+0.01}_{-0.00}$ &  \\ 
HD 269953 & $5843^{+52}_{-46}$ & $0.35^{+0.14}_{-0.15}$ & $493^{+6}_{-6}$ & $5.407^{+0.022}_{-0.022}$ & $4.10^{+0.14}_{-0.13}$ & $0.16^{+0.02}_{-0.02}$ &  \\ 
CPD-69 496 & $5995^{+72}_{-36}$ & $0.49^{+0.06}_{-0.20}$ & $267^{+4}_{-3}$ & $4.919^{+0.030}_{-0.019}$ & $4.02^{+0.18}_{-0.06}$ & $0.12^{+0.03}_{-0.02}$ &  \\ 
HD 268819 & $6090^{+20}_{-17}$ & $1.10^{+0.09}_{-0.09}$ & $352^{+2}_{-2}$ & $5.185^{+0.006}_{-0.003}$ & $3.43^{+0.09}_{-0.09}$ & $0.00^{+0.00}_{-0.00}$ &  \\ 
HD 268687 & $6166^{+69}_{-61}$ & $2.98^{+0.13}_{-0.13}$ & $257^{+2}_{-2}$ & $4.934^{+0.020}_{-0.020}$ & $1.57^{+0.12}_{-0.12}$ & $0.03^{+0.02}_{-0.02}$ &  \\ 
HD 270050 & $6447^{+62}_{-53}$ & $0.59^{+0.13}_{-0.07}$ & $217^{+3}_{-2}$ & $4.867^{+0.023}_{-0.020}$ & $4.03^{+0.06}_{-0.12}$ & $0.03^{+0.02}_{-0.02}$ &  \\ 
HD 269697 & $6564^{+44}_{-44}$ & $1.41^{+0.09}_{-0.11}$ & $335^{+1}_{-1}$ & $5.274^{+0.013}_{-0.014}$ & $3.25^{+0.10}_{-0.09}$ & $0.08^{+0.01}_{-0.01}$ &  \\ 
HD 269902 & $6586^{+26}_{-24}$ & $3.99^{+0.01}_{-0.01}$ & $281^{+2}_{-2}$ & $5.128^{+0.003}_{-0.003}$ & $0.68^{+0.01}_{-0.01}$ & $0.00^{+0.00}_{-0.00}$ &  \\ 
HD 269331 & $6685^{+33}_{-28}$ & $2.44^{+0.10}_{-0.10}$ & $264^{+2}_{-2}$ & $5.098^{+0.007}_{-0.004}$ & $2.26^{+0.10}_{-0.10}$ & $0.00^{+0.00}_{-0.00}$ &  \\ 
HD 269840 & $6713^{+55}_{-53}$ & $1.68^{+0.10}_{-0.10}$ & $332^{+1}_{-1}$ & $5.304^{+0.014}_{-0.014}$ & $3.02^{+0.09}_{-0.09}$ & $0.18^{+0.01}_{-0.01}$ &  \\ 
CD-69 310 & $6869^{+65}_{-55}$ & $2.66^{+0.11}_{-0.13}$ & $186^{+1}_{-1}$ & $4.841^{+0.018}_{-0.016}$ & $2.09^{+0.12}_{-0.11}$ & $0.02^{+0.01}_{-0.01}$ &  \\ 
HD 269982 & $6906^{+61}_{-54}$ & $2.40^{+0.12}_{-0.11}$ & $228^{+2}_{-1}$ & $5.029^{+0.016}_{-0.015}$ & $2.35^{+0.10}_{-0.11}$ & $0.08^{+0.01}_{-0.01}$ &  \\ 
HD 269857 & $6962^{+78}_{-59}$ & $1.98^{+0.15}_{-0.12}$ & $291^{+2}_{-2}$ & $5.253^{+0.020}_{-0.016}$ & $2.78^{+0.11}_{-0.13}$ & $0.13^{+0.01}_{-0.01}$ &  \\ 
HD 269651 & $7288^{+49}_{-51}$ & $2.99^{+0.01}_{-0.01}$ & $136^{+2}_{-2}$ & $4.673^{+0.004}_{-0.004}$ & $1.85^{+0.02}_{-0.01}$ & $0.00^{+0.00}_{-0.00}$ &  \\ 
HD 269662 & $7327^{+53}_{-49}$ & $3.99^{+0.01}_{-0.01}$ & $153^{+2}_{-2}$ & $4.786^{+0.004}_{-0.004}$ & $0.86^{+0.02}_{-0.01}$ & $0.00^{+0.00}_{-0.00}$ &  \\ 
HD 33579 & $7398^{+59}_{-54}$ & $2.99^{+0.00}_{-0.01}$ & $352^{+5}_{-5}$ & $5.525^{+0.004}_{-0.004}$ & $1.88^{+0.02}_{-0.01}$ & $0.00^{+0.00}_{-0.00}$ &  \\ 
HD 269807 & $7492^{+7}_{-267}$ & $0.51^{+1.31}_{-0.01}$ & $159^{+2}_{-1}$ & $4.848^{+0.007}_{-0.048}$ & $4.38^{+0.01}_{-1.37}$ & $0.04^{+0.01}_{-0.03}$ &  \\ 
HD 269604 & $7806^{+38}_{-38}$ & $3.98^{+0.02}_{-0.04}$ & $138^{+1}_{-1}$ & $4.804^{+0.003}_{-0.003}$ & $0.99^{+0.04}_{-0.02}$ & $0.00^{+0.00}_{-0.00}$ &  \\ 
HD 269762 & $8248^{+2}_{-3}$ & $1.01^{+0.01}_{-0.00}$ & $112^{+1}_{-1}$ & $4.720^{+0.006}_{-0.006}$ & $4.05^{+0.00}_{-0.01}$ & $0.00^{+0.00}_{-0.00}$ &  \\ 
HD 269661 & $10493^{+6}_{-12}$ & $2.01^{+0.01}_{-0.01}$ & $122^{+1}_{-1}$ & $5.207^{+0.010}_{-0.011}$ & $3.47^{+0.01}_{-0.01}$ & $0.13^{+0.01}_{-0.01}$ & \checkmark \\ 
HD 268946 & $10494^{+5}_{-10}$ & $2.01^{+0.01}_{-0.00}$ & $159^{+2}_{-2}$ & $5.442^{+0.009}_{-0.009}$ & $3.47^{+0.00}_{-0.01}$ & $0.11^{+0.01}_{-0.01}$ & \checkmark \\ 
SK -69 99 & $10494^{+4}_{-9}$ & $2.01^{+0.01}_{-0.00}$ & $67^{+1}_{-1}$ & $4.689^{+0.007}_{-0.008}$ & $3.47^{+0.00}_{-0.01}$ & $0.08^{+0.01}_{-0.01}$ & \checkmark \\ 
HD 268971 & $10494^{+4}_{-9}$ & $2.01^{+0.01}_{-0.00}$ & $75^{+1}_{-1}$ & $4.790^{+0.008}_{-0.009}$ & $3.47^{+0.00}_{-0.01}$ & $0.02^{+0.01}_{-0.01}$ & \checkmark \\ 
HD 269781 & $10494^{+4}_{-9}$ & $2.01^{+0.01}_{-0.00}$ & $184^{+2}_{-2}$ & $5.568^{+0.008}_{-0.008}$ & $3.47^{+0.00}_{-0.01}$ & $0.14^{+0.01}_{-0.01}$ & \checkmark \\ 
HD 268727 & $10494^{+4}_{-9}$ & $2.00^{+0.01}_{-0.00}$ & $87^{+0}_{-0}$ & $4.920^{+0.004}_{-0.004}$ & $3.47^{+0.00}_{-0.01}$ & $0.15^{+0.00}_{-0.00}$ & \checkmark \\ 
HD 269787 & $10495^{+4}_{-8}$ & $2.00^{+0.01}_{-0.00}$ & $139^{+1}_{-1}$ & $5.322^{+0.006}_{-0.006}$ & $3.47^{+0.00}_{-0.01}$ & $0.16^{+0.00}_{-0.00}$ & \checkmark \\ 
HD 268949 & $11738^{+9}_{-20}$ & $2.01^{+0.01}_{-0.00}$ & $53^{+0}_{-0}$ & $4.686^{+0.007}_{-0.007}$ & $3.66^{+0.01}_{-0.01}$ & $0.10^{+0.01}_{-0.01}$ & \checkmark \\ 
\enddata
\end{deluxetable*}

\subsection{Atmospheric Parameters}\label{subsec:atmospheric}

The bulk of our analysis is focused on the data products produced by the MIKE pipeline. However, for the template-fitting procedure adopted by \citetalias{chen24}, we applied the following corrections. After reduction, we downloaded literature radial velocities for our targets from SIMBAD\footnote{\url{https://simbad.u-strasbg.fr/simbad/}} and used the {\tt noao.rv.rvcorrect} and {\tt noao.imred.echelle.dopbor} tasks within IRAF \citep{tody86,tody93} to correct our spectra to the rest frame. We then flux calibrated each individual order using the flux standards we observed as well as the {\tt standard}, {\tt sensfunc}, and {\tt calibrate} tasks within the IRAF {\tt noao.imred.echelle} package. Finally, individual orders were stitched together using {\tt noao.imred.echelle.scombine}. The resulting flux-calibrated spectra were then fit by \citetalias{chen24} with ATLAS9 models \citep{castelli03}, from which they measured effective temperatures, surface gravities, spectroscopic luminosities (defined as $\mathcal{L}\propto T_{\rm eff}^4/g$), and extinctions. Because the data were flux calibrated, the distance to the LMC is known, and the ATLAS9 models are given in units of surface flux, this procedure also yielded measurements of the photospheric radius (via the scaling from the model to the data) and bolometric luminosity (via the Stefan-Boltzmann law).

As discussed in \S\ref{sec:intro}, many of our surface gravity measurements appear to be biased upward by unaccounted-for broadening sources. In Appendix \ref{app:updatedparams}, we present an update to the procedure presented by \citetalias{chen24} that includes our inferred broadening parameters as well as an improved treatment of the errors on the flux measurements. However, as in \citetalias{chen24}, our errors are still underestimated; while this does not impact our downstream analysis, the reader is encouraged to take this into account when using our measurements. The resulting parameters, presented in Table \ref{tab:atmoparam}, are used throughout the rest of this work. 

We note that, while the majority of our sample have spectroscopic parameters that are in line with typical YSGs (and are consistent with their literature spectral types), eight stars had derived effective temperatures at the hot edge of the prior --- i.e., $\teff>4$ --- corresponding to spectral types earlier than A0, implying that these are in fact late B supergiants. This is somewhat surprising given their literature spectral types (all A0), and inclusion in the sample of LMC YSGs in \citet{neugent12_ysg}. Of course, \citeauthor{neugent12_ysg} used 2MASS photometry to estimate effective temperatures, which is effective for YSGs but loses sensitivity for earlier A stars \citep[e.g.][]{dornwallenstein23}. Furthermore, at these high temperatures, a large amount of the flux is at the bluest end of the MIKE wavelength coverage, where the throughput rapidly drops off. As a result, it is likely that our flux calibration and spectral order stitching procedure introduced some amount of bias into the final spectra of these stars, making our effective temperatures for these stars unreliable. For this reason, as well as to ensure that our sample is focused exclusively on the YSG phases of stellar evolution, we exclude these stars from further analysis, leaving a final sample of 32 YSGs.

\section{Line Profile Measurements}\label{sec:ACID}

The profile of an individual atomic absorption line is dependent on the properties of the atom and the specific electronic transition, as well as the structure of the photosphere in which the line formed. This structure also informs various broadening processes that impact the line. Most importantly for this analysis, these broadening processes include the macroscale velocity fields within the outermost layers of the star being observed \citep{gray05}. Indeed, line profiles in cool supergiants can be used to place powerful constraints on the dynamics in the photosphere \citep[e.g.][]{lopezariste25}.

Because individual absorption lines may only be deep enough to probe the center of the underlying profile, combining multiple line profiles (and correcting for line-to-line variations in depth) can yield a much higher signal-to-noise measurement, especially in the wings of the profile. The cross-correlation method has been used widely, especially in the context of using the derived profile for precision radial velocity measurements \citep[e.g.][]{blancocuaresma14,blancucuaresma19}. However, the cross-correlation method can produce spurious line profiles or even fail when lines are blended, as in low-resolution spectra or spectra with significant broadening. The Least-Squares Deconvolution (LSD) technique \citep{donati97} mitigates these effects. However, it requires spectra that have been normalized; i.e., the continuum flux level is one. Continuum normalization is a subtle and difficult art \citep[e.g.][]{gunn83}. This is particularly true of high-resolution echelle spectra, especially those containing broad features which may take up a significant fraction of an individual spectral order, as is the case for these data. 

\citet{dolan24} introduced the Accurate Continuum fItting and Deconvolution (ACID) method, a variant on LSD that, among other improvements, simultaneously fits the line profile and the continuum. Using the Python implementation of ACID\footnote{\url{https://acid-code.readthedocs.io/en/latest/index.html}}, we derive line profiles for each order of each spectrum. We provide ACID with the wavelength and flux for each order, in addition to the errors reported by the pipeline, removing any pixels with 0, negative, or {\tt NaN} flux. We also provide a linelist for each star from the Vienna Atomic Line Database \citep{ryabchikova97,ryabchikova15,kupka99,kupka00,pakhomov19}. Linelists are generated using the {\sc Extract Stellar} function, assuming atmospheric parameters from \citetalias{chen24}, a microturbulent velocity of 10 km s$^{-1}$ (broadly suitable for YSGs --- e.g., \citealt{nieuwenhuijzen12} --- and consistent with the results of this study), and a minimum line depth of 0.007. We note that, despite the fact that the surface gravities in \citetalias{chen24} are slightly overestimated in some cases, \citet{dolan24} demonstrated that ACID's procedure for masking bad lines makes it robust against moderate mismatches between the data and the linelist.\footnote{If our newly-derived effective temperatures were significantly different from those in \citetalias{chen24} such that the linelists were completely inapplicable, then a different procedure would be warranted. In theory, we could perform an iterative process, deriving line profiles, fitting them as in \S\ref{subsec:profilefit}, remeasuring atmospheric parameters using updated broadening parameters as in Appendix \ref{app:updatedparams}, downloading new linelists and repeating until both the profiles and the stellar parameters converged. However, such a procedure would be computationally expensive, and only result in minor improvement to the line profiles.}

The line profile and associated errors are computed on a grid of velocities centered on the literature radial velocity for each star, extending for $\pm200$ km s$^{-1}$. To compute the grid spacing, we use the {\tt calc\_deltav} function within ACID, which uses the wavelength spacing of the data to compute an appropriate velocity spacing. To ensure all orders are calculated on the same velocity grid, we use the smallest computed velocity spacing for all orders. The continuum is fit with a 9$^{th}$-order polynomial. Lines with more than a $2\sigma$ discrepancy between the ACID model and the data are removed (the default is $1\sigma$). Telluric lines are drawn from a telluric linelist provided in the {\sc iSpec} package \citep{blancocuaresma14,blancucuaresma19}, and 3\AA~on either side of each telluric line with a depth greater than 0.1 is masked out. Additionally, we mask out 3\AA~on either side of the first seven Balmer lines, the Na doublet, and the Ca II infrared triplet (after redshifting those wavelengths to the restframe of the spectrum). These lines are excluded as they are both broad enough that they deviate from the shape of the majority of the rest of the atomic lines, and in some cases (Balmer lines, Ca II triplet) are observed in emission. 

Because of the spectral types of our targets, the bluest orders of the MIKE spectra frequently did not have enough signal to robustly measure line profiles, and ACID would fail. For these orders, we record the profile as equal to 1 for all velocity pixels, with an error of $\infty$. As a result, such a profile has zero weight in the profile-combining procedure we describe below.\footnote{We note that the choice of ACID parameters described above results in a very low signal-to-noise profile derived from the spectrum of HD 269661, which was observed during a brief period of notably bad seeing. We were able to successfully measure line profiles for this star by modifying our procedure to fit the continuum with a 5$^{th}$-order polynomial, only masking out lines with a $1\sigma$ discrepancy from the ACID model, and masking any points that deviated from the median of each order by more than 5 standard deviations. HD 269661 was still ultimately excluded from further analysis due to its high effective temperature.} 

To combine the line profiles for each order, we perform a least-squares fit to each profile with a linear continuum plus a Gaussian feature, and divide the profiles by the inferred continuum to ensure the continuum level is at one for each order's profile. For each order, we then compute the first velocity moment (see \S\ref{subsec:moments}) to identify and correct any order-to-order variations in the wavelength solution that impact the centroids of the inferred profile. Finally, we compute the average profile in each velocity bin, weighing by the inverse of the squared error in that pixel; as described above, orders on which ACID failed were assigned an error of $\infty$, resulting in a weight of 0 in this weighted averaging.

Importantly, we note that the resulting line profiles are the ``typical'' line profile for each star. They are derived using individual atomic lines that are formed in varying depths throughout the photosphere. This procedure is necessary due to how broadened (and thus blended) the individual lines are, and the range of spectral types contained in our sample, which make it impossible to, for example, only look at the profile of one suitably strong line. However, this choice results in a loss of any sensitivity to depth-dependent variations in the photospheric velocity field. In principal, we could use linelists separated by formation depth, and perform a tomographic analysis. Indeed, this method has already been developed for asymptotic giant branch stars \citep{alvarez01} and applied successfully to red supergiants using extremely high signal to noise data \citep[e.g.][]{josselin07,kravchenko19,jadlovsky24}, and would be an excellent subject for future investigations of YSGs observed at higher SNR.

\subsection{Moment Analysis}\label{subsec:moments}

\begin{deluxetable*}{lllr}
\tabletypesize{\scriptsize}
\tablecaption{The first three statistical moments of our measured line profiles.\label{tab:moments}}
\tablehead{\colhead{Common Name} & \colhead{$\vfirst$} & \colhead{$\vsecond$} & \colhead{$\vthird$}\\
\colhead{} & \colhead{[km s$^{-1}$]} & \colhead{[$10^2$(km s$^{-1}$)$^2$]} & \colhead{[$10^2$(km s$^{-1}$)$^3$]}} 
\startdata
 HV 883 & $277.66\pm0.22$ & $3.8\pm0.1$ & $110\pm18$ \\ 
{[W60]} D17 & $317.58\pm0.17$ & $5.6\pm0.1$ & $-176\pm14$ \\ 
2MASS J05344326-6704104 & $305.84\pm0.18$ & $3.9\pm0.1$ & $-124\pm15$ \\ 
SP77 31-16 & $250.61\pm0.13$ & $5.1\pm0.1$ & $-95\pm11$ \\ 
SK -69 148 & $279.14\pm0.16$ & $4.9\pm0.1$ & $-34\pm10$ \\ 
RM 1-77 & $249.59\pm0.18$ & $5.3\pm0.1$ & $-275\pm17$ \\ 
SP77 48-6 & $238.44\pm0.13$ & $4.0\pm0.0$ & $-36\pm8$ \\ 
HD 269723 & $308.41\pm0.16$ & $7.4\pm0.1$ & $229\pm14$ \\ 
 HV 2450 & $293.02\pm0.60$ & $8.9\pm0.3$ & $-491\pm48$ \\ 
HD 269879 & $306.98\pm0.23$ & $4.4\pm0.0$ & $9\pm6$ \\ 
HD 269110 & $233.97\pm0.19$ & $6.9\pm0.1$ & $11\pm18$ \\ 
HD 269070 & $252.17\pm0.14$ & $5.6\pm0.1$ & $-22\pm13$ \\ 
HD 268828 & $253.82\pm0.19$ & $4.5\pm0.1$ & $-237\pm17$ \\ 
HD 268865 & $280.41\pm0.23$ & $3.1\pm0.0$ & $71\pm9$ \\ 
HD 269953 & $249.80\pm0.11$ & $7.0\pm0.0$ & $54\pm7$ \\ 
CPD-69 496 & $250.48\pm0.15$ & $2.8\pm0.0$ & $-15\pm2$ \\ 
HD 268819 & $252.38\pm0.13$ & $6.0\pm0.0$ & $-95\pm5$ \\ 
HD 268687 & $255.46\pm0.19$ & $5.6\pm0.1$ & $-23\pm10$ \\ 
HD 270050 & $304.34\pm0.17$ & $4.0\pm0.0$ & $27\pm3$ \\ 
HD 269697 & $302.41\pm0.15$ & $4.6\pm0.0$ & $-41\pm7$ \\ 
HD 269902 & $267.21\pm0.79$ & $34.3\pm0.5$ & $656\pm72$ \\ 
HD 269331 & $257.87\pm0.17$ & $9.9\pm0.1$ & $13\pm8$ \\ 
HD 269840 & $301.78\pm0.20$ & $5.3\pm0.1$ & $-72\pm10$ \\ 
CD-69 310 & $269.29\pm0.71$ & $11.2\pm0.2$ & $-134\pm19$ \\ 
HD 269982 & $262.44\pm0.27$ & $9.8\pm0.1$ & $114\pm7$ \\ 
HD 269857 & $274.05\pm0.20$ & $7.1\pm0.0$ & $54\pm4$ \\ 
HD 269651 & $276.04\pm0.50$ & $13.5\pm0.2$ & $10\pm27$ \\ 
HD 269662 & $269.56\pm0.49$ & $17.2\pm0.2$ & $288\pm35$ \\ 
HD 33579 & $265.41\pm0.18$ & $12.0\pm0.1$ & $71\pm7$ \\ 
HD 269807 & $310.31\pm0.31$ & $5.4\pm0.1$ & $149\pm10$ \\ 
HD 269604 & $280.10\pm0.53$ & $13.2\pm0.1$ & $115\pm21$ \\ 
HD 269762 & $286.30\pm0.55$ & $26.1\pm0.3$ & $986\pm48$ \\ 
\enddata
\end{deluxetable*}

In order to characterize the shapes of our line profiles, we measure the statistical moments of each line profile. Following \citet{balona86}, we compute the first, second, and third velocity moments, $\vfirst$, $\vsecond$, and $\vthird$, incorporating Sheppard's correction for binned data \citep{sheppard1897}. For readers unfamiliar with statistical moments, $\vfirst$ is an average velocity (i.e., the systemic radial velocity), $\vsecond$ gives a sense of the square of the width of the profile (i.e., the scale of velocity deviations in the photosphere), and $\vthird$ gives the skewness (i.e., profiles with positive/negative values appear deeper towards positive/negative velocities). Mathematical derivations of these quantities can be found in \citet{balona86}. We tabulate these quantities and their associated errors in Table \ref{tab:moments}. For the remaining analysis on the line profiles, we center the profiles on $\vfirst$. 

\subsection{Profile Fitting}\label{subsec:profilefit}

In addition to a moment-based characterization of the line profiles, we also fit our profiles with various mixtures of broadening kernels. We first consider radial-tangential macroturbulence following \citet{gray05}, defining the macroturbulent profile as
\begin{multline}
    \mathcal{F}_{RT}(\Delta v) = \pi\bigg[\frac{2A_{R}\Delta v}{\pi^{1/2}\zeta_R^2}\int_0^{\zeta_R/\Delta v}e^{-1/u^2}du + \\ \frac{2A_{T}\Delta v}{\pi^{1/2}\zeta_T^2}\int_0^{\zeta_T/\Delta v}e^{-1/u^2}du\bigg] 
\end{multline}
where we set the macroturbulent velocities $\zeta_R = \zeta_T = \vturb$ and amplitudes $A_R = A_T = A_{RT}$.\footnote{This is a standard choice when fitting line profiles with a macroturbulent component \citep[e.g.][]{simon-diaz14,simon-diaz17}. Nominally, it is possible to disentangle the radial and tangential components of the macroturbulent velocity field. However, such work is difficult in all but the highest signal-to-noise regime \citep{serebiakova23}, which is not possible in the present work.}

Additionally, we consider a Gaussian function:
\begin{equation}
    \mathcal{F}_{\rm mic} = \frac{1}{\sqrt{2\pi\vmicro^2}}e^{-\Big(\frac{\Delta v^2}{2\vmicro^2}\Big)}
\end{equation}
whose integral (i.e., equivalent width) is unity, and whose width is set by $\vmicro$, where we are deliberately making an analogy with the microturbulent broadening parameter, $\xi$. We stress that $\vmicro$ is {\it not} strictly the microturbulent velocity. Rather, this Gaussian component corresponds to a combination of microturbulence and the unknown underlying thermal profile width $kT/m$, such that $\vmicro^2 = (kT/m )^2 + \xi^2$. However, the thermal profiles of supergiants are notably narrow \citep{maury97}, while microturbulence in YSGs is $\sim$10 km s$^{-1}$ \citep{nieuwenhuijzen12}, implying that any Gaussian component to our line profiles is dominated by microturbulence, and so we will refer to this component as the microturbulent component hereafter.

Finally, we also consider rotational broadening (including linear limb darkening) as in e.g. \citet{simon-diaz17}, again following \citealt{gray05}:
\begin{equation}
    \Theta_{\rm rot} = \frac{2(1-\epsilon)\sqrt{1-\big(\frac{\Delta v}{\vsini}\big)^2}+\frac{1}{2}\pi\epsilon\Big(1-\big(\frac{\Delta v}{\vsini}\big)^2\Big)}{\pi\vsini(1-\epsilon/3)}
\end{equation}
with the projected equatorial rotational velocity $\vsini$ and the linear limb darkening coefficient $\epsilon$.

\begin{figure*}[t!]
\centering
\plottwo{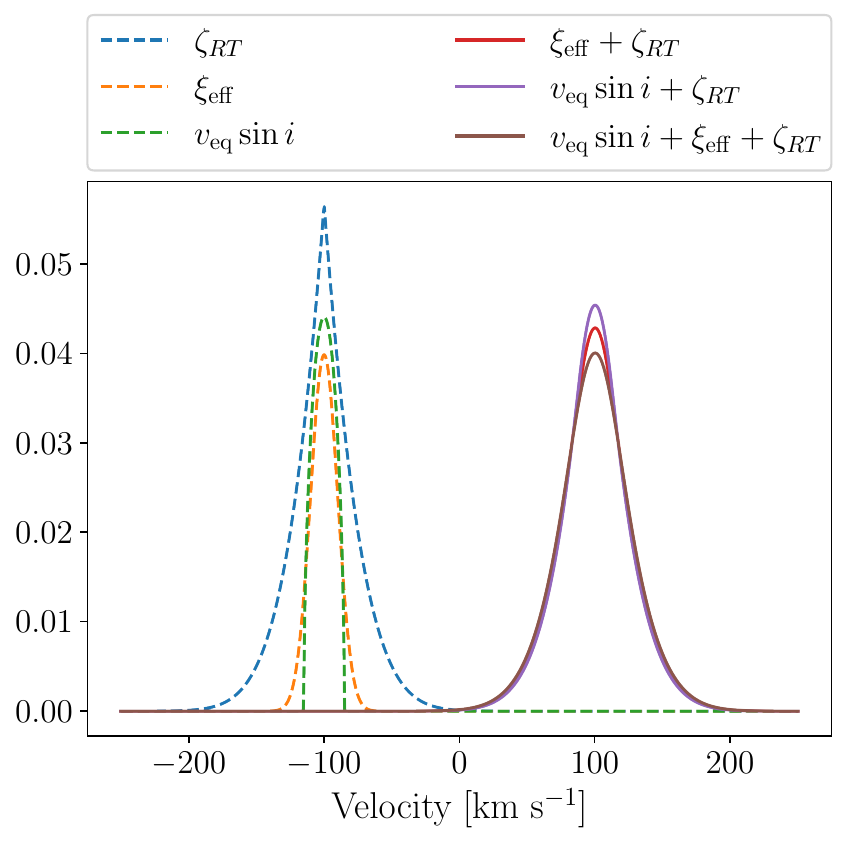}{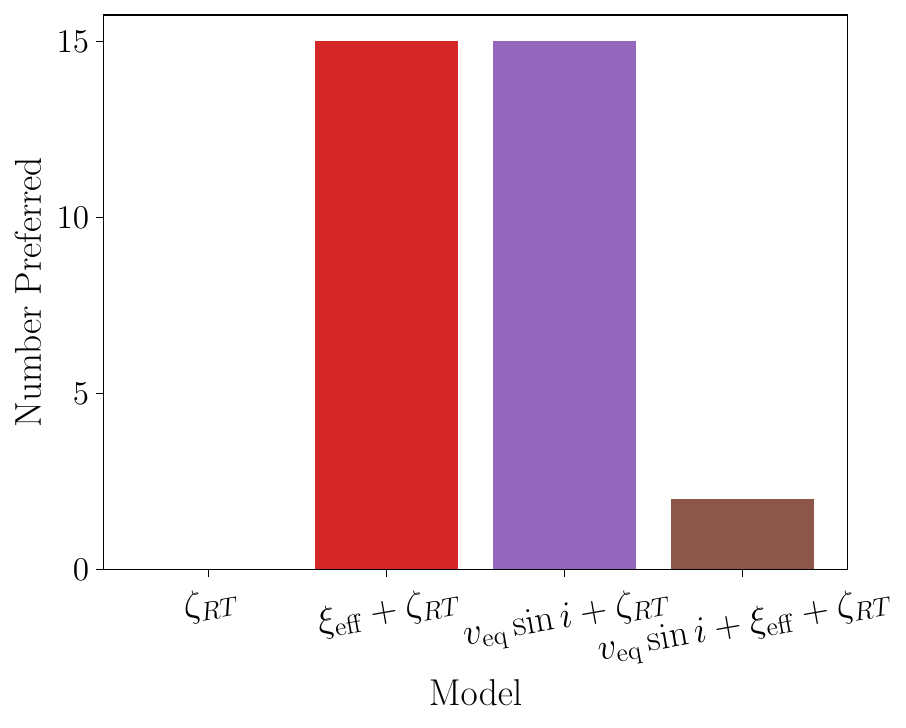}
\caption{{\it Left}: A comparison of our line profile models, each denoted by their corresponding velocity. The individual broadening kernels are shown as dashed lines to the left, while the convolved models are shown as solid lines to the right. {\it Right}: The number of stars for each a given model yields the minimum $BIC$. While the line profiles are all highly broadened by macroturbulence, we find that an additional component is necessary to fit all of the observed line profiles.}\label{fig:model_selection}
\end{figure*}

Using these three broadening kernels, we consider four different models: one with macroturbulence alone, one with a convolution of $\mathcal{F}_{RT}$ and $\mathcal{F}_{\rm mic}$, one with a convolution of $\mathcal{F}_{RT}$ and $\Theta_{\rm rot}$, and one with a convolution of all three kernels. We illustrate the shapes of our models in the left panel of Figure \ref{fig:model_selection}. We show the three individual broadening kernels as dashed lines to the left of the plot and the convolved kernels as solid lines to the right. Each individual model is labeled by the corresponding velocity/velocities. For this illustration, we adopt $\vturb=50$ km s$^{-1}$, $A_{RT}=0.4$, $\vmicro=10$ km s$^{-1}$, $\vsini=15$ km s$^{-1}$, and $\epsilon=0.3$. 

We note that because all four models contain the amplitude parameter $A_{RT}$, there is no need to further scale the microturbulent or rotational kernels, which are already normalized. Any additional normalization factors would be perfectly degenerate with $A_{RT}$. Finally, we introduce a wavelength shift, $v_0$, to account for offsets in how the observed profiles were centered. To fully resolve the shapes of each broadening component, we computed the profiles and convolution on a grid of 500 evenly-spaced velocity points before interpolating back down to the resolution of the data.

We now define a likelihood function for each model:
\begin{equation}\label{eq:prof_ll}
    \ln \mathcal{L} = -\frac{1}{2}\sum_i \frac{(p_i-\hat{p}_i)^2}{s_i^2} + \ln (2\pi s_{i}^2)
\end{equation}
where $p_i$ is the observed profile in the $i^{th}$ velocity bin, $\hat{p}_i$ is the model, and 
\begin{equation}
    s_i^2 = \sigma_i^2 + f^2\hat{p}_i^2
\end{equation}
is an error term that includes an additional parameter, $f$, which represents a fractional error relative to the model by which the errors on our derived profiles are underestimated. Using {\sc scipy}, we identify the parameters for each model and for each observed line profile that numerically maximize\footnote{In practice, we minimize the negative of Eq. \eqref{eq:prof_ll}} Eq. \eqref{eq:prof_ll}. Using the maximum likelihood solution, we then calculate the Bayesian Information Content \citep[$BIC$][]{schwarz78} of each model:
\begin{equation}
    BIC = -2\ln\mathcal{L}_{\rm max} + k\ln N
\end{equation}
where $\ln\mathcal{L}_{\rm max}$ is the maximized log-likelihood, $k$ is the number of free parameters in the model (4, 5, 6, and 7 for the macroturbulence, macroturbulence $+$ microturbulence, macroturbulence $+$ rotation, and macroturbulence $+$ microturbulence $+$ rotation models respectively), and $N$ is the number of points in the observed line profile. Then, for each star's line profile, we identify the model whose maximum likelihood parameters minimize the $BIC$.

The right panel of Figure \ref{fig:model_selection} shows the number of stars for which each model we consider minimizes the $BIC$. While all of the observed line profiles are strongly broadened by macroturbulence, none of the observed line profiles are best-fit by $\mathcal{F}_{RT}$ alone, indicating that a narrow component to the model is statistically required. However, a model containing all three of the broadening kernels described above is, in most cases, too complex, and does not yield a sufficient increase in log-likelihood to justify the increased model complexity. Ultimately, 30 out of 32 stars are best-fit (in the minimum-$BIC$ sense) by macroturbulence and either Gaussian microturbulence or rotation. 

We pause here to note that, in the OB star literature, studies have found correlations between measured macroturbulent and rotational velocities. For example, Figure 4 of \citet{simon-diaz17} shows measurements of $\vturb$ (which they denote $v_{\rm mac}$) and $\vsini$. To test for a correlation in their data, we use a Spearman rank correlation coefficient \citep{spearman04}, which is a nonparametric test of the existence of a monotonic relationship between two quantities. Applying a two-sided Spearman test to the stars in their sample with $\vturb$ measurements (as opposed to upper limits), we find a Spearman coefficient of $R=0.72$, and $p\ll0.01$; the two quantities are correlated with an extreme degree of statistical significance. 

This finding is puzzling given the inclination-dependence of the rotational velocity measurement, which is thus a lower limit on the true rotational velocity. If a correlation between said true rotational velocity and macroturbulence existed, then one would expect the observed trend to actually be an envelope or boundary, with $\vsini$ falling below the underlying correlation. Of course, much work has already been done attempting to characterize the biases of line broadening analyses, particularly those done in the Fourier domain \citep[e.g.][]{aerts14,serebiakova23}. However, as our community continues to push the limits of both sample size and signal-to-noise, we are presented with new opportunities to reevaluate the physical and statistical justifications for our model choices, particularly when inferring rotational velocities in the presence of other, comparably strong sources of line broadening. For example, \citet{markova25} demonstrated that neglecting to include microturbulence when measuring macroturbulent and rotational velocities can bias these measurements upwards.

Ultimately, we adopt the macroturbulence $+$ microturbulence model for physical reasons. YSGs on their first crossing of the HR diagram may still be rotating around $\sim$10 km s$^{-1}$, at least while in the faintest and hottest --- and therefore most compact --- regions of the part of the HR diagram containing our sample. However, when we experimented with fitting a macroturbulence $+$ rotation model, the inferred equatorial rotation speeds exceed these values in many cases, including the coolest and most luminous --- and therefore largest --- stars. Furthermore, we found correlations between $\vturb$ and $\vsini$ in this experiment. If the spin axes of the stars in our sample are randomly oriented --- and there is no reason to think they would not be --- then both the magnitude of the inferred $\vsini$ values, and the presence of this correlation would be physically improbable. 

Of course, the binary fraction in massive stars is high \citep[e.g.][]{sana12,demink13,moe17}. It is possible that, in binary systems, orbital angular momentum may be transferred to the surface via tidal effects, depending on the mass ratio and orbital separation. However, because these stars are radially extended, they have already expanded and spun down by necessity. Furthermore, tides are inefficient at angular momentum transport \citep{vick21,mackleod22}, and the spin-up timescale is quite long. As a result, only binaries in extremely tight orbits could spin up a YSG primary via tides. Simultaneously, binaries with {\it too} tight an orbit would have merged as the primary expanded off the main sequence. Therefore, only a very narrow range of initial orbital periods would produce just the right conditions to spin up a YSG via tides such that the observed (projected) rotational velocities could be comparable to the macroturbulent velocity \citep{eldridge17}. It would require an unrealistic degree of fine-tuning for this to have occured for all YSGs in our sample. 

We note that a future analysis may be able to separate the narrow broadening component into microturbulence and some additional contribution from rotation, but, as our $BIC$ analysis implies, such an analysis would require much higher signal-to-noise line profiles, or the introduction of additional data such as photometric variability. 
With sufficiently high-resolution observations and the ability to resolve individual lines, tomographic measurements may further aid in disentangling the effects of convection and rotation, which may vary differently with depth (see, e.g. \citealt{kravchenko18,kravchenko19,kravchenko21} and a recent review by \citealt{chiavassa24}). However, we anticipate that disentangling convection from rotation via tomography will be especially difficult in supergiant stars, where large-scale individual convective plumes span a large range in optical depth \citep{brun09,goldberg22} and can resemble rotation even on large scales when the atmosphere is insufficiently spatially-resolved \citep{kervella18,ma24}.
We note also that in cases where rotation is favored by the BIC, the rotational broadening is always at least a factor of two narrower than the macroturbulent broadening.

With our chosen line profile model, we then wish to estimate and infer confidence intervals for each of the model parameters in a Bayesian setting, with a focus on constraining the turbulent velocities $\vturb$ and $\vmicro$. We define a prior such that $v_0 / {\rm km \ s^{-1}} \sim \mathcal{N}(0.0,5.0)$, $A_{RT}\sim \mathcal{HN}(10)$, $\vturb / {\rm km \ s^{-1}} \sim \mathcal{HN}(100)$, $\vmicro / {\rm km \ s^{-1}} \sim \mathcal{HN}(100)$, and $\log f\sim\mathcal{U}(-10,-1)$, where $\mathcal{N(\mu,\sigma)}$ is the normal distribution with mean $\mu$ and variance $\sigma^2$, $\mathcal{U}(l,h)$ is the uniform distribution between $l$ and $h$, and $\mathcal{HN}(\sigma)$ is a half-normal distribution, centered on 0, with variance $\sigma^2$. 

With our likelihood from Eq. \eqref{eq:prof_ll} and priors, we use {\sc emcee} \citep{foremanmackey13} to perform Markov Chain Monte Carlo (MCMC) to estimate the posterior probability distribution of each parameter. We initialize 32 walkers in a tight Gaussian sphere centered on the maximum-likelihood parameters. To ensure convergence, every 100 steps, we use {\sc emcee} to estimate the autocorrelation time for each parameter, averaged across Markov chains. We continue taking steps until the estimated autocorrelation times have converged, and we have taken more samples than 100 times the longest autocorrelation time --- typically $\mathcal{O}(10^4)$ steps. We then discard the first two autocorrelation times worth of steps, so that the remaining samples are samples from the genuine posterior probability distribution. From these samples, we take the median value as our measurement for each parameter, and use the 16$^{th}$ and 84$^{th}$ percentile values to define a confidence interval. We tabulate our measurements of $\vturb$ and $\vmicro$ in Table \ref{tab:vturb_vmicro}.

We note that, during the initial ``burn-in'' phase of parameter sampling, sometimes the value $\vmicro$ would become small enough that the corresponding profile is not resolved (i.e., the array containing the profile was all zeros), and the convolution of the two profiles becomes numerically unstable. In these cases, we simply return the macroturbulent profile. This only occurred for small numbers of samples, and none of our measured velocity values fall below this resolution limit. 

\begin{deluxetable}{lll}
\tabletypesize{\scriptsize}
\tablecaption{Measurements of $\vturb$ and $\vmicro$ for the stars in our sample.\label{tab:vturb_vmicro}}
\tablehead{\colhead{Common Name} & \colhead{$\vturb$} & \colhead{$\vmicro$}\\
\colhead{} & \colhead{[km s$^{-1}$]} & \colhead{[km s$^{-1}$]}} 
\startdata
 HV 883 & $20.2_{-0.6}^{+0.6}$ & $6.8_{-0.3}^{+0.3}$ \\ 
{[W60]} D17 & $25.5_{-0.6}^{+0.5}$ & $9.4_{-0.2}^{+0.2}$ \\ 
2MASS J05344326-6704104 & $19.6_{-0.4}^{+0.3}$ & $6.2_{-0.2}^{+0.2}$ \\ 
SP77 31-16 & $29.9_{-0.6}^{+0.6}$ & $10.8_{-0.2}^{+0.3}$ \\ 
SK -69 148 & $27.6_{-1.8}^{+1.5}$ & $10.9_{-0.6}^{+0.7}$ \\ 
RM 1-77 & $19.2_{-0.3}^{+0.3}$ & $6.2_{-0.2}^{+0.2}$ \\ 
SP77 48-6 & $19.0_{-5.0}^{+2.4}$ & $14.1_{-0.7}^{+1.2}$ \\ 
HD 269723 & $41.9_{-1.0}^{+0.9}$ & $13.2_{-0.4}^{+0.5}$ \\ 
 HV 2450 & $26.8_{-0.8}^{+0.7}$ & $6.1_{-0.4}^{+0.4}$ \\ 
HD 269879 & $32.6_{-0.7}^{+0.7}$ & $8.6_{-0.3}^{+0.4}$ \\ 
HD 269110 & $28.9_{-3.2}^{+2.3}$ & $14.7_{-0.9}^{+1.1}$ \\ 
HD 269070 & $32.0_{-0.8}^{+0.7}$ & $11.4_{-0.3}^{+0.3}$ \\ 
HD 268828 & $26.0_{-0.4}^{+0.4}$ & $6.9_{-0.2}^{+0.2}$ \\ 
HD 268865 & $28.4_{-0.4}^{+0.4}$ & $5.8_{-0.2}^{+0.2}$ \\ 
HD 269953 & $36.5_{-1.4}^{+1.3}$ & $17.2_{-0.5}^{+0.5}$ \\ 
CPD-69 496 & $27.9_{-0.6}^{+0.5}$ & $8.0_{-0.3}^{+0.3}$ \\ 
HD 268819 & $39.8_{-1.0}^{+1.0}$ & $11.1_{-0.5}^{+0.5}$ \\ 
HD 268687 & $37.6_{-0.7}^{+0.7}$ & $11.4_{-0.3}^{+0.3}$ \\ 
HD 270050 & $32.5_{-0.4}^{+0.4}$ & $8.1_{-0.2}^{+0.2}$ \\ 
HD 269697 & $34.1_{-0.3}^{+0.3}$ & $8.6_{-0.2}^{+0.2}$ \\ 
HD 269902 & $70.4_{-22.1}^{+10.2}$ & $35.0_{-3.8}^{+6.7}$ \\ 
HD 269331 & $50.7_{-1.0}^{+0.9}$ & $12.4_{-0.5}^{+0.5}$ \\ 
HD 269840 & $39.0_{-0.6}^{+0.6}$ & $9.5_{-0.3}^{+0.3}$ \\ 
CD-69 310 & $44.0_{-1.0}^{+1.0}$ & $5.6_{-0.6}^{+0.6}$ \\ 
HD 269982 & $50.0_{-0.8}^{+0.8}$ & $7.8_{-0.5}^{+0.5}$ \\ 
HD 269857 & $42.5_{-0.8}^{+0.7}$ & $8.8_{-0.4}^{+0.4}$ \\ 
HD 269651 & $41.0_{-4.7}^{+3.4}$ & $17.4_{-1.4}^{+1.7}$ \\ 
HD 269662 & $36.8_{-23.2}^{+16.1}$ & $26.5_{-5.6}^{+4.5}$ \\ 
HD 33579 & $46.8_{-2.0}^{+1.8}$ & $18.1_{-0.8}^{+0.8}$ \\ 
HD 269807 & $30.4_{-0.9}^{+0.9}$ & $8.2_{-0.4}^{+0.5}$ \\ 
HD 269604 & $58.4_{-1.1}^{+1.1}$ & $6.8_{-0.7}^{+0.7}$ \\ 
HD 269762 & $82.0_{-1.3}^{+1.3}$ & $11.5_{-0.8}^{+0.8}$ \\ 
\enddata
\end{deluxetable}

\begin{figure*}[t!]
\plottwo{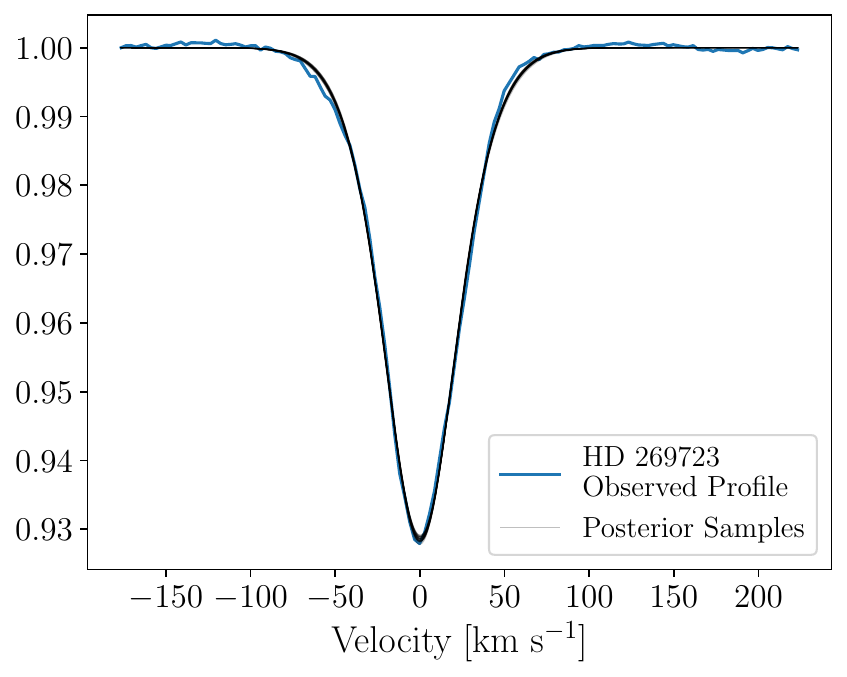}{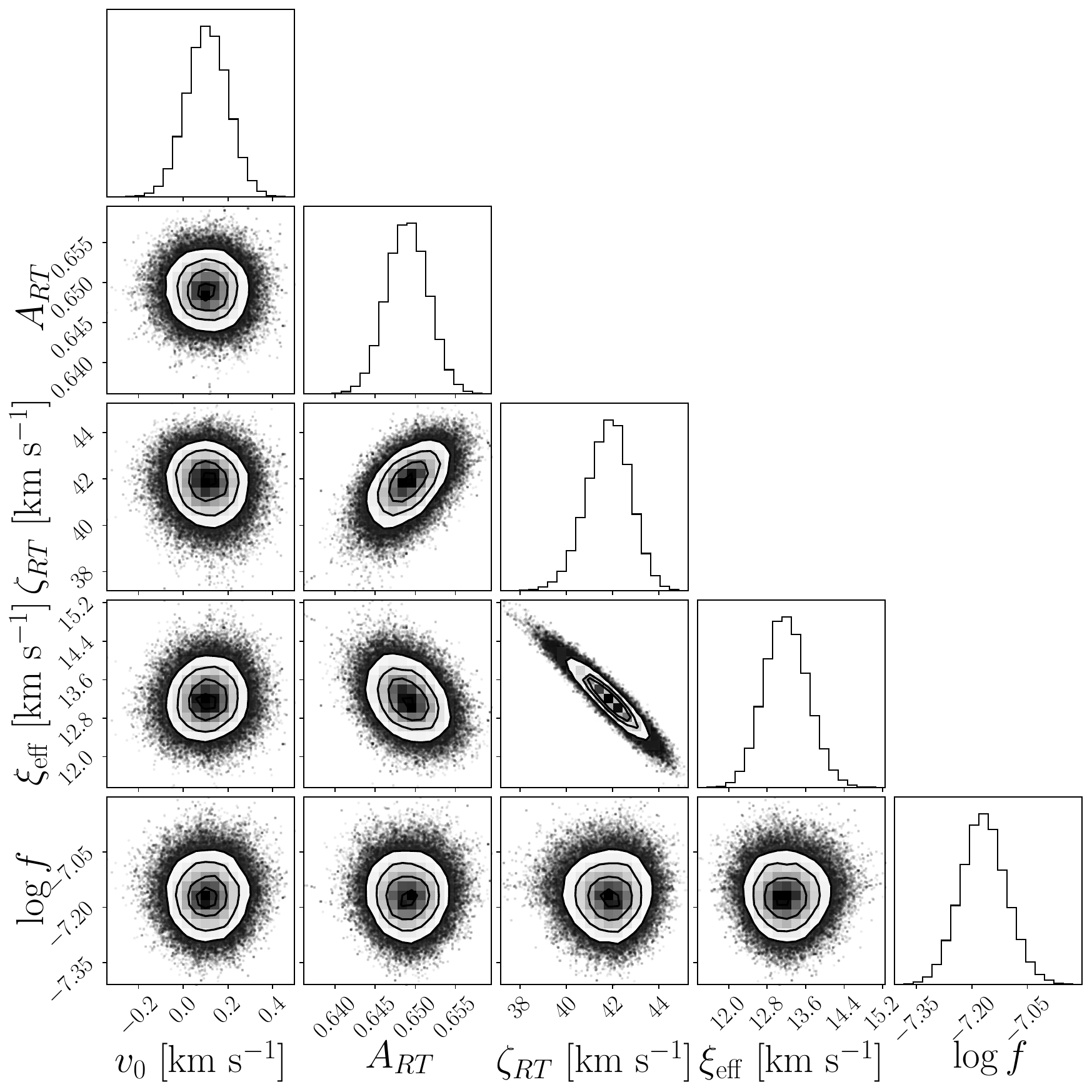}
\caption{Example results from our line profile fitting procedure. ({\it Left}): The observed line profile of HD 269723 is shown in blue, with 32 samples from the posterior probability distribution of our model parameters shown in black. The profile is well-fit by our model. ({\it Right}): Samples from the posterior probability distribution. Panels along the diagonal show the marginalized posterior probability distributions for $v_0$, $A_{RT}$, $\vturb$, $\vmicro$, and $\log f$. Off-diagonal panels show the joint posterior probability distributions for combinations of these variables.}\label{fig:example_fit}
\end{figure*}

The left panel of Figure \ref{fig:example_fit} shows the line profile for HD 269723 in blue. Black lines are 32 random draws from the posterior samples, shown in the space of the data. The right panel of the Figure contains a corner plot illustrating the marginalized posterior probability distributions of the model parameters. We derive tight constraints on all parameters, and find a covariance between $\vturb$ and $\vmicro$.

\begin{figure*}[t!]
\centering
\includegraphics[width=0.9\textwidth]{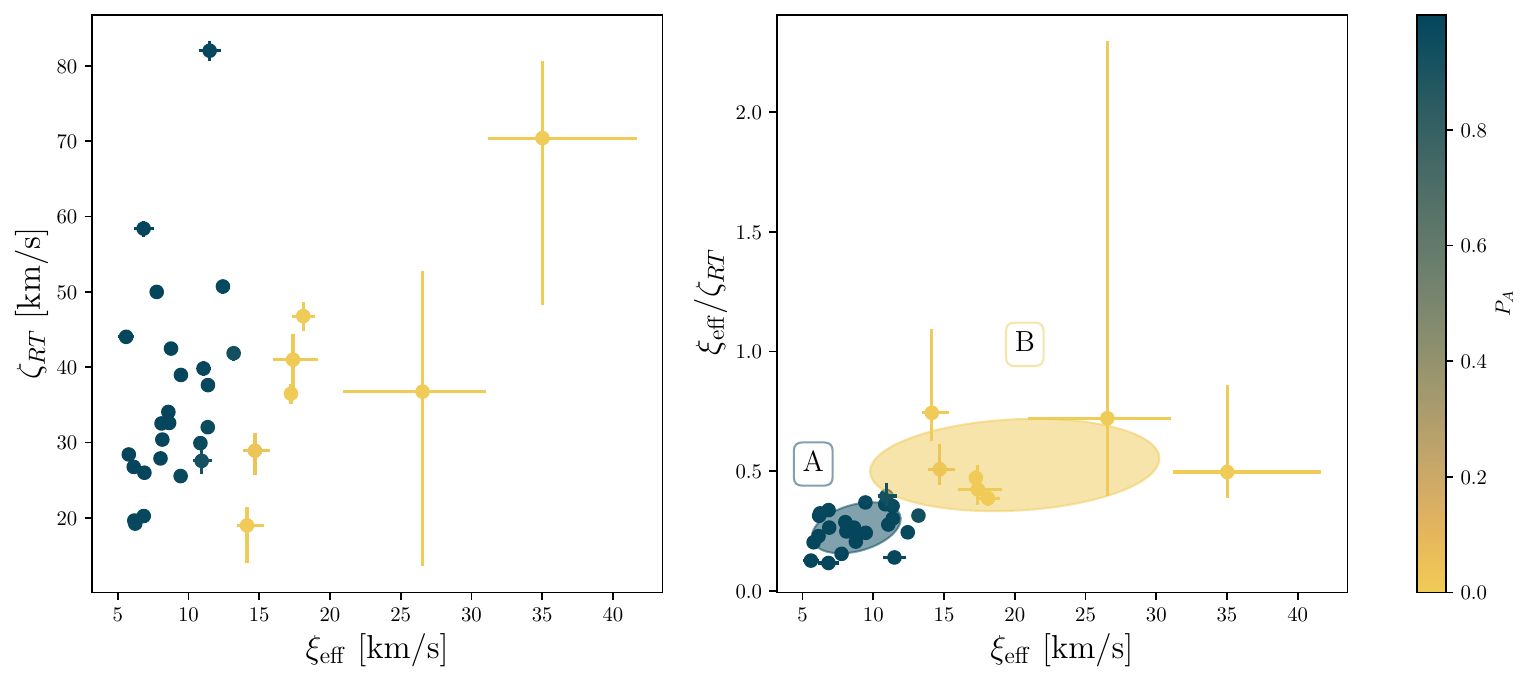}
\caption{{\it Left}: $\vturb$ versus $\vmicro$. The majority of the stars have $\vmicro\lesssim10$ km s$^{1}$, consistent with past studies of YSGs \citep[e.g.][]{kovtyukh12,nieuwenhuijzen12}. However, some of the stars exhibit much higher microturbulent velocities, and appear to be separated from the main locus of YSGs. {\it Right}: Our measurements recast into $\vmicro/\vturb$ versus $\vmicro$, in which the separation between the two groups can be seen, which we denote group A and group B. Ellipses indicate the mean and covariance of the Gaussian Mixture Model components used to separate the two groups. Points in both panels are colored by the probability of belonging to group A.}\label{fig:turb_vels_GMM}
\end{figure*}

\section{Results}\label{sec:results}

\subsection{Turbulence}\label{subsec:turbrot}

We plot $\vturb$ versus $\vmicro$ in the left hand panel of Figure \ref{fig:turb_vels_GMM}. For all stars, $\vturb > \vmicro$, indicating that macroturbulent velocity is the dominant source of line broadening in our sample of YSGs. Indeed, while our measurements of $\vturb$ range from $\approx$20 to $\approx$130 km s${-1}$, the majority of our $\vmicro$ measurements are around 10 km s$^{-1}$ or less, in agreement with observational studies of microturbulence in YSGs \citep{kovtyukh12,nieuwenhuijzen12,markova25}. This finding lends further credence to our physical argument for choosing Gaussian microturbulence over rotation for the narrow broadening component in the case of YSGs.

Examining Figure \ref{fig:turb_vels_GMM} further, a small number of YSGs appear to have $\vmicro$ values above the typical value of 10 km s$^{-1}$. These stars appear to be separated from the main group of YSGs in the sample, with the dividing line being at roughly 13-14 km s$^{-1}$. The difference between these two subsets of stars can be seen more clearly when we plot $\vmicro/\vturb$\footnote{We compute this ratio directly from the posterior samples of each profile fit (instead of taking the ratio of the median parameter estimates) in order to derive robust estimates and uncertainties.} versus $\vmicro$ in the right hand panel of Figure \ref{fig:turb_vels_GMM}. In this space, the stars are neatly separated into two groups: group A, which captures the YSGs with ``typical'' microturbulent velocities, and group B, which captures the stars with higher $\vmicro$ values at a given $\vturb$. Text boxes in the Figure show which group is which. We note that this finding is not due to, for example, our macroturbulence $+$ microturbulence model being a poor fit for the group B stars. There is, in fact, no relationship between the stars in group B, and the stars which were better fit (in a $BIC$ sense) by a different model.

To characterize these groups, we fit the data in the right hand panel of Figure \ref{fig:turb_vels_GMM} with a two-component Gaussian mixture model (GMM) using {\sc scikit-learn}. The GMM can predict the probability of each point belonging to group A, $P_A$\footnote{the probability of belonging to group B, is simply $P_B = 1-P_A$}, which we use to color each point in both panels of Figure \ref{fig:turb_vels_GMM}. When we do this, we find that the $P_A$ values are closely clustered around 0 or 1, i.e., the model finds that the two groups are well-separated. To determine if our choice of only two components is justified, we use a ``leave one out'' cross-validation procedure. Given 32 data points, we fit each possible subset containing 31 points with a GMM, using from one to six components. We find that two components yields the highest per-sample log-likelihood averaged over subsets of the data. To illustrate the individual components of the GMM, we use shaded ellipses generated using the means the covariances of the Gaussian model components. 

In addition to the probability that a given star belongs to a given group, the GMM also yields weights, indicating the overall probability of any star in our sample belonging to a group. With these weights, we estimate that 77.2\% of YSGs belong to group A, and 22.8\% belong to group B. Taking these these groups as illustrative, we can compare distributions, similarities, and differences as a function of other physical and observational properties.

\subsubsection{Turbulence as a Function of Stellar Parameters}\label{subsubsec:atmotrends}

\begin{figure*}[t!]
\centering
\includegraphics[width=0.9\textwidth]{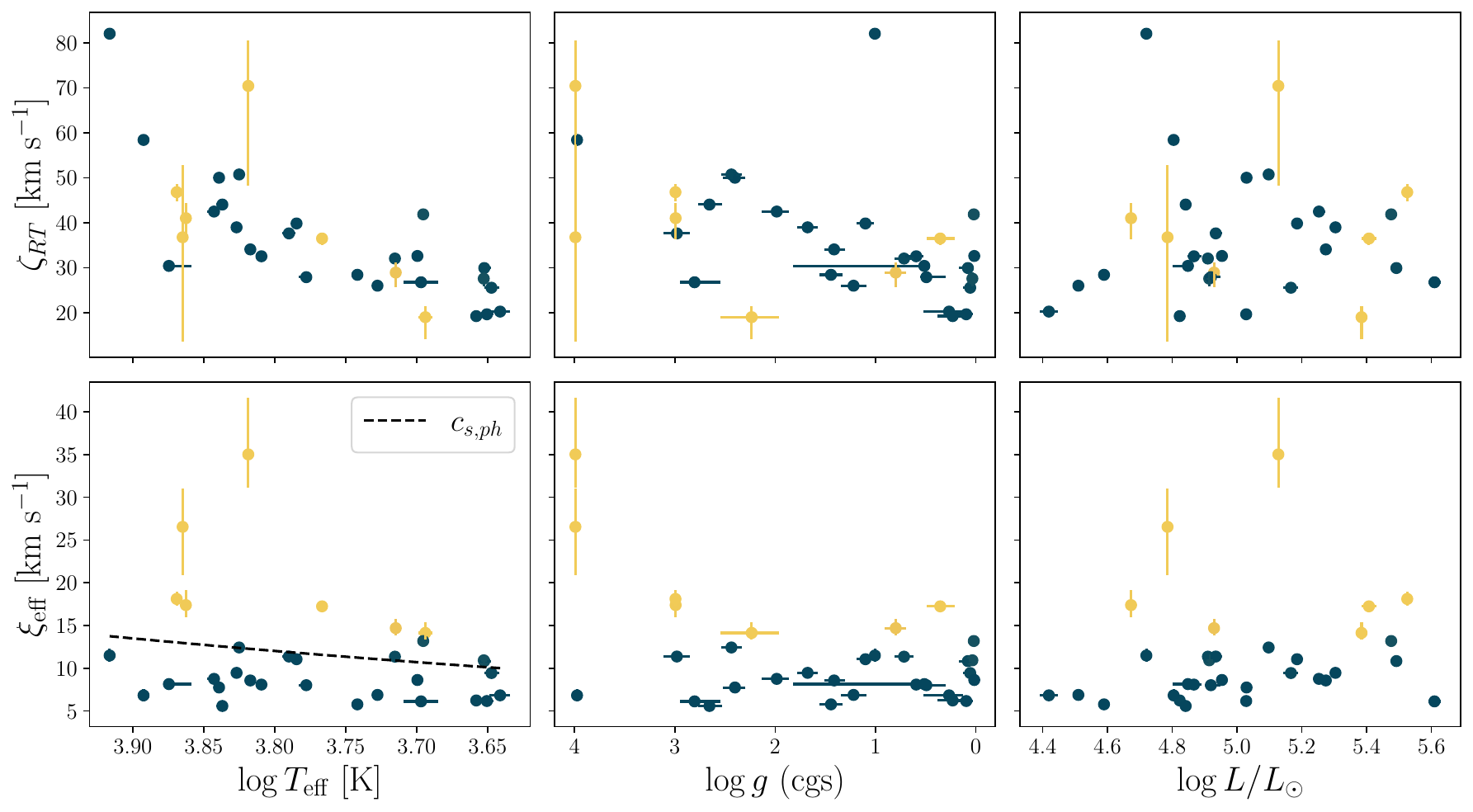}
\caption{{\it Top}: $\vturb$ as a function of $\teff$ (top-left panel), $\log g$ (top-center panel), and $\lum$ (top-right panel), colored by $P_A$ as in Figure \ref{fig:turb_vels_GMM}. {\it Bottom}: Similar for $\vmicro$ as a function of atmospheric parameters. Additionally, we show the photospheric sound speed for an ideal gas as a function of effective temperature in the lower-left hand panel as a dashed black line.}\label{fig:atmo_turb}
\end{figure*}

The top row of panels in Figure \ref{fig:atmo_turb} shows $\vturb$ as a function of our newly-measured stellar parameters, with points colored by $P_A$ as defined above. Similarly, the bottom panels show $\vmicro$ as a function of stellar properties. We find that $\vturb$ is positively correlated with both effective temperature and surface gravity, while $\vmicro$ is positively correlated with $\lum$. These findings are in agreement with past studies of macroturbulence in the massive star regime \citep{simon-diaz17,deburgos24}, as well as in lower mass cool stars \citep{gray05}. However, we don't see any obvious trends between macroturbulence and luminosity, nor between microturbulence and both temperature and surface gravity. We find no extremely strong differences between groups A and B as a function of luminosity, gravity, and temperature, except that for the group B stars, there does appear to be a sharp increase in microturbulence at higher temperatures and higher surface gravities, but with such few stars, it is difficult to determine the statistical significance of this result. 

We do pause to note that this increase occurs around $\teff\approx3.85$, which is the same temperature as the ``yellow void,'' the poorly-populated region of the HR diagram where stars' envelopes become dynamically unstable. However, the group B stars that are the most obvious outliers in this panel have luminosities below $\lum=5.2$, while the yellow void is only observed above $\lum\approx5.4$ \citep{dejager98}. While this observation may ultimately be a coincidence, such little is understood about YSGs compared to their red and blue cousins to make it worth mentioning.

Finally, we note that, for an ideal gas, the sound speed in the photosphere is
\begin{equation}
c_{s,ph} = \sqrt{\frac{\gamma k_B T_{\rm eff}}{\mu m_p}}
\end{equation}
given the adiabatic index, $\gamma$, and mean molecular weight $\mu$. We compute $c_{s,ph}$, assuming $\gamma=5/3$, $\mu=0.6$, for the range of temperatures spanning our sample, which we plot in the lower left hand panel of Figure \ref{fig:atmo_turb} as a dashed black line. Critically, for typical YSG effective temperatures, the ideal gas sound speed is 10-15 km s$^{-1}$, which is roughly equal to the dividing line between groups A and B. This could be just an interesting coincidence, but we can indulge in some speculation: the macroturbulent velocities of our YSGs are all above the sound speed by at least a factor of two. This finding is consistent with the fact that, in cool supergiants, it is expected that the convection is near- or super-sonic \citep[e.g.][]{josselin07}, especially so in very luminous supergiants \citep[e.g.][]{jiang15,jiang18,goldberg22,schultz2023}. Additionally, it implies that whichever physical process underlies macroturbulence is out of causal contact with the microturbulent component.  Furthermore, in the group B stars, the microturbulence is also above the calculated $c_{s,ph}$ scale. While our calculation of $c_{s,ph}$ is hardly a robust estimate of the true sound speed in the photosphere of a realistic YSG, this implies that turbulence on both large and small size scales could drive shocks in YSG atmospheres, which may contribute to the driving mechanism of mass loss in this poorly-studied evolutionary phase.

\subsection{Turbulence and Line Asymmetry}

\begin{figure}[t!]
\centering
\includegraphics[width=0.4\textwidth]{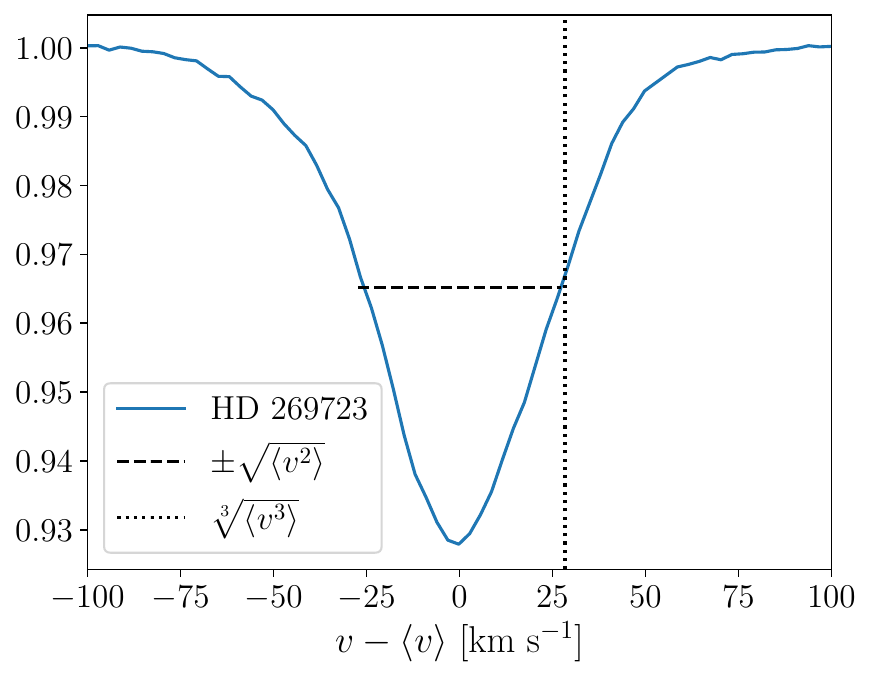}
\caption{Line profile for HD 269723, shifted by the first velocity moment, and expanded to highlight the asymmetry of the profile. The black vertical dotted line shows the cube-root of the third velocity moment.}\label{fig:example_assymetric}
\end{figure}

Our derived line profiles are notably asymmetric, as indicated by our measured values of the profile skewness, $\vthird$. As an example, Figure \ref{fig:example_assymetric} shows the line profile for HD 269723, centered on the first velocity moment, $\vfirst$, as in Figure \ref{fig:example_fit}. However, here we zoom in to the core of the profile to highlight the asymmetry. The black horizontal dashed line shows the spread of velocities corresponding to $\pm\sqrt{\vsecond}=\pm27.3$ km s$^{-1}$. The black vertical dotted line shows the velocity corresponding to $\sqrt[3]{\vthird}= 28.4$ km s${-1}$. The line is notably wider on the red (positive velocity) side than the blue (negative velocity) side. Turbulence on spatial scales much smaller than the size of the visible disk of the star, i.e., the stellar radius, cannot cause line asymmetry as our measurements are disk-averaged and have no spatial information. Therefore, line asymmetry, along with our $\vturb$ measurements, is a direct probe of the velocities of large-scale flows of material. Notably, in OB stars where $\vturb$ is the dominant contribution to line broadening, there is a positive correlation between $\vturb$ and the relative skewness $R_{sk}$, which is defined as
\begin{equation}
    R_{sk} = \frac{\vthird}{\vsecond^{\frac{3}{2}}}
\end{equation}
in the sense that stars with higher turbulent velocities have higher absolute $R_{sk}$ values \citep{simon-diaz17}. 

\begin{figure*}[t!]
\centering
\includegraphics[width=0.9\textwidth]{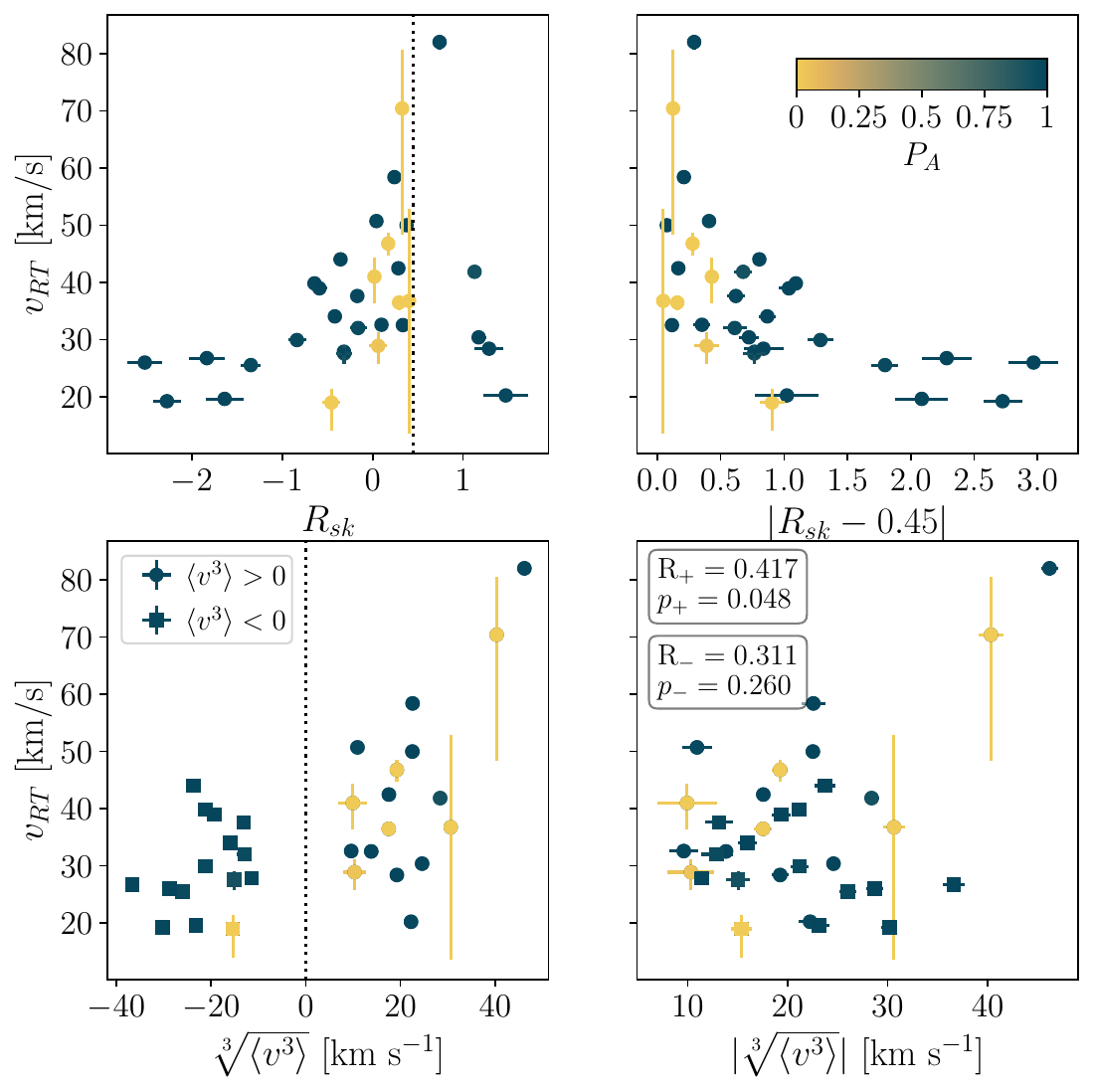}
\caption{{\it Top Left}: $\vturb$ as a function of $R_{sk}$, with points colored by $P_A$ as in Figure \ref{fig:turb_vels_GMM}. The points are peaked around $R_{sk}=0.45$, shown as the dotted vertical black line. {\it Top Right}: $\vturb$ as a function of $|R_{sk}-0.45|$, illustrating the anti-correlation between the two quantities. {\it Bottom} $\vturb$ as a function of the third velocity moment (bottom left) and absolute value of the third velocity moment (bottom right). Points with positive skew are shown as circles, and points with negative skew are shown as squares; the line where the skew is equal to zero is shown as a dotted vertical black line. The Spearman $R$ coefficient and corresponding $p$-value for the correlation between the two quantities are shown in text boxes, with $+/-$ subscripts indicating the values for the positively/negatively skewed stars respectively. A statistically significant correlation is only detected for the positively skewed stars.}\label{fig:vturb_rsk}
\end{figure*}

Using our line moment measurements, we compute the relative skewness, and plot macroturbulent velocity as a function of $R_{sk}$ in the top left panel of Figure \ref{fig:vturb_rsk}, with points colored by $P_A$. We find that, in direct contrast with \citeauthor{simon-diaz17}, the quantities are anti-correlated: stars with higher $\vturb$ have values of $R_{sk}$ closer to zero. However, the stars with the highest $\vturb$ values do not have $R_{sk}=0$, and instead cluster around $R_{sk}\approx0.45$, which we show as a vertical dotted black line. This ``zero point offset'' may be a systematic effect due to an asymmetry in the instrumental line profile, which is not accounted for in either the MIKE reduction pipeline or our line profile measurement procedure. Furthermore, we find that the group B YSGs have smaller values of $R_{sk}$ than the group A stars. To examine this further, the top right panel of Figure \ref{fig:vturb_rsk} shows $\vturb$ as a function of absolute $R_{sk}$ after subtracting an offset of 0.45. This panel shows both that group B YSGs have smaller $R_{sk}$, and that we find an anti-correlation between $\vturb$ and line profile asymmetry as captured by $R_{sk}$ that directly contrasts the results from similar studies in OB stars. 

We now focus further on the finding that stars in group B have smaller $R_{sk}$ values. One way of statistically testing this finding is via tests on the equality of the variances of $R_{sk}$ between the two subsamples; i.e., if the group B absolute $R_{sk}$ values are smaller than in group A, we would expect the variance of these measurements to be smaller. The classic method of testing equalities of variances is an F-test: we compute $F$, the ratio of the variances of $R_{sk}$ in the two groups, which, under the null hypothesis, is F-distributed with degrees of freedom $\nu_A$ and $\nu_B$ equal to one less than the number of stars in each group. We find $F=15.285$ for degrees of freedom $\nu_A=24$, $\nu_B=6$, which corresponds to a $p$-value of 0.001, which we use to reject the null hypothesis.

Notably an F-test is only valid when the two samples are normally distributed. We test the  normality of the $R_{sk}$ values of each sample using a Shapiro-Wilkes test \citep{shapiro65}. We find $p=0.412$ for group A, and $p=0.198$ for group B: i.e., there is insufficient statistical evidence for deviations from normality in either sample's $R_{sk}$ values. Of course, the sample sizes are small, especially for group B. An alternative test for the equality of variances in samples that are not normally distributed is Bartlett's test \citep{bartlett37}, which we perform using {\sc scipy}, and which yields a $p$-value of 0.003, which we use to again reject the null hypothesis. In short: the line profiles of YSGs in group B have smaller $R_{sk}$ values.

However, this is not the complete story; the bottom panels of Figure \ref{fig:vturb_rsk} show $\vturb$ as a function of $\sqrt[3]{\vthird}$ (bottom left) and $|\sqrt[3]{\vthird}|$ (bottom right), illustrating that the stars with the highest macroturbulent velocities have {\it more} assymetric profiles as captured by the third velocity moment/skewness. However, the growth of $\vthird$ with $\vturb$ is ``slower'' than the growth of $\vsecond$ with $\vturb$ (which, unsurprisingly, is well-fit by a second order polynomial, i.e., $\vsecond \propto \vturb^2$), which is why we find the opposite trend in relative skewness.

Furthermore, there also appears to be some difference between the stars with positive skewness (circular points in the lower panels of Figure \ref{fig:vturb_rsk}) and those with negative skewness (square points); we show where $\sqrt[3]{\vthird}=0$ km s$^{-1}$ with a dashed vertical black line to denote the separation between the two. For the stars with positive skew, the two quantities appear to be positively correlated. However, no such trend is apparent in the stars with negative skew. Indeed, there are no negatively skewed stars with $\vturb$ values above 45 km s${-1}$. This may imply physical differences between the relative velocity scales, amplitudes, and surface area coverage of upward-flowing (i.e., negative radial velocity) versus downward-flowing (positive radial velocity) material in the turbulent flow. 

To quantify the apparent correlation between $\vturb$ and $\vthird$, we use a Spearman rank correlation coefficient \citep{spearman04}, which is a nonparametric test of the existence of a monotonic relationship between two quantities. We compute the Spearman $R$ coefficient between $\vturb$ and $\sqrt[3]{\vthird}$ using the {\tt spearmanr} function within {\sc scipy}, which we perform on the positively and negatively skewed stars separately. This test also yields a $p$-value, indicating the statistical significance of the relationship. For the positively-skewed stars, we specifically test whether the correlation between skew and $\vturb$ is positive, while for the negatively-skewed stars we use a two-sided test (i.e., testing for any correlation at all). In the bottom right panel of Figure \ref{fig:vturb_rsk}, we report the resulting $R$ and $p$ values in text boxes. We indicate the values for the positively/negatively skewed stars with $+/-$ subscripts respectively. We detect a positive correlation for the positively skewed stars with $p=0.048$, and no such correlation for the negatively skewed stars. We note that, while $p=0.048$ satisfies a $p<0.05$ significance test, it only barely does so. Choosing to instead perform a two-sided Spearman test yielded a statistically insignificant result ($p=0.096$). Future work will be required to verify the existence and physical origin of this correlation. It is nonetheless interesting that all of the stars in our sample have a minimum amount of asymmetry, with $|\vthird|\gtrsim10$ km s${-1}$. This indicates a clear deviation from small-scale isotropic turbulence.If pulsations were the only driving factor of these asymmetries, then we would expect to observe at least some of our sample at a phase with zero asymmetry. This is perhaps not surprising, as the morphology of convection contains asymmetries betweens lower-density upflows and denser downflows, even in the sun \citep[e.g.][]{stein89}, though the scale of the asymmetry here is larger. 

\begin{figure*}[t!]
\centering
\includegraphics[width=\textwidth]{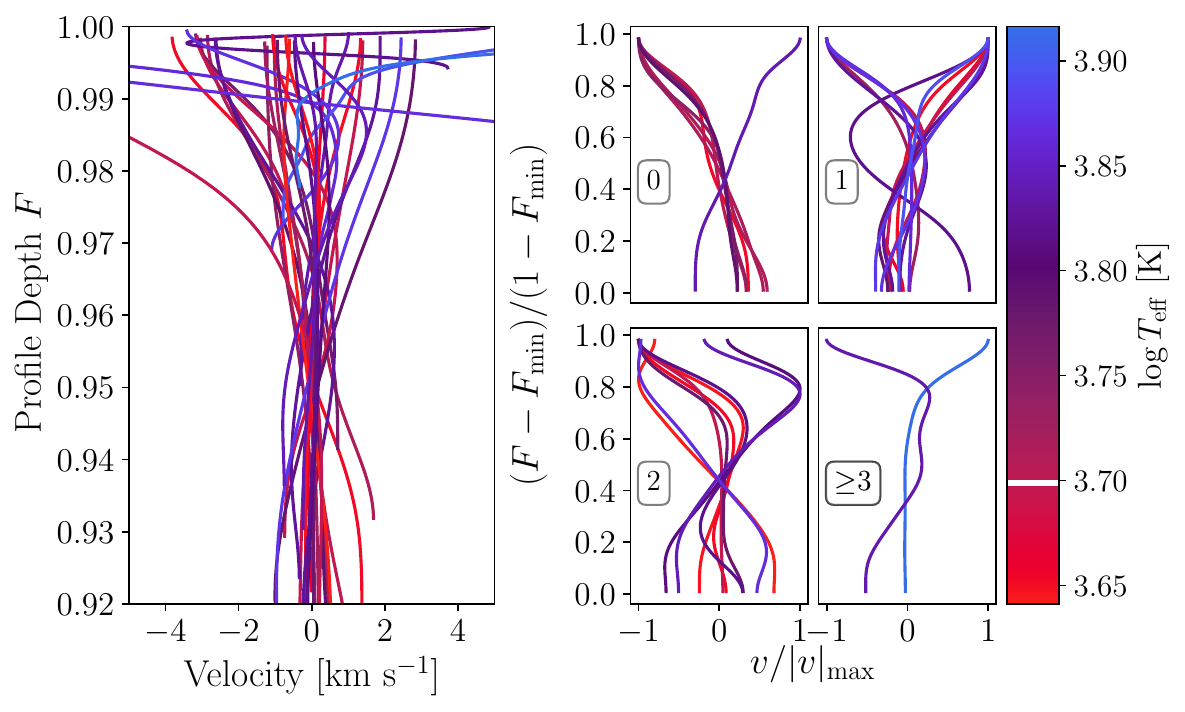}
\caption{{\it Left}: Bisectors computed from our line profiles, color-coded by effective temperature as shown by the colorbar. The white line in the colorbar shows where $\teff=3.7$, corresponding to a spectral type of G2 \citep{tabernero18}. {\it Right}: Bisectors, sorted by the number of extrema from zero (top left), to three or more (bottom right), and scaled such that the bottom of each bisector is at 0, and the maximum horizontal excursion of each bisector is $\pm1$. The number of extrema is indicated in a text box in each panel.}\label{fig:bisectors_Next}
\end{figure*}

\subsubsection{Higher Order Asymmetry Probed By Line Bisectors}

As a further probe of line asymmetry, we also compute the bisectors of each line profile; i.e., as a function of depth, we take the average velocity of the two points in the profile closest to that depth. While perfectly symmetric profiles will have a vertical bisector (the profile is identical to either side at each depth, so the average is zero), deviations in the bisectors from zero are thus indicative of velocity structures in the photosphere that vary with physical depth, as the core of the line is formed in higher photospheric layers than the wings. 

While the vast majority of bisector analyses are done in the Sun (for which suitable signal to noise and spectral resolution is easily obtainable), a number of studies have looked at bisectors in cool supergiants. A classic work is \citet{gray86}, who found that Ib supergiants with spectral types G2 and later showed C-shaped bisectors, as might be expected for stars with granulation, as upward flowing material in granulation cells is hotter, and thus brighter, causing a blueward shift in core of the line, and a redward shift in the wings. However, stars with earlier spectral types displayed backward C-shaped bisectors. Adopting the effective temperature scale from \citet{tabernero18}, this divergence in behavior occurs at 5000 K, i.e., $\teff\approx3.7$. \citeauthor{gray86} proposed that this inversion of the bisector results from much faster moving upward-welling material being significantly brighter (i.e. hotter) than material falling back down. Alternatively, they proposed an ``expanding star'' hypothesis (i.e. mass loss), which they discarded on the basis of the resulting mass loss rates being far too high given the understanding of mass loss in cool supergiants at the time.

To compute our bisectors, we first shift and scale our line profiles such that the minimum flux is always 0, which allows us to uniformly compare bisectors between stars. Because our profiles represent the line depth on a regular grid of velocities, we then linearly interpolate them onto a regular grid of {\it depths}, from 0.02 to 1.0 with a spacing of 0.01. Finally, we use a 10 pixel Gaussian smooth along the depth axis to reduce the impact of noise in the profiles before computing the line bisectors.

We find an astonishing range of behavior. The left panel of Figure \ref{fig:bisectors_Next} shows our bisectors, color-coded by effective temperature as indicated by the colorbar, with redder bisectors corresponding to cooler YSGs. We indicate where $\teff=3.7$ (i.e., G2 where the two bisector behaviors found by \citealt{gray86} diverge) as a white line in the colorbar. The ``classic'' supergiant bisectors presented by \citet{gray86} show one or no extrema. By contrast, many of our bisectors show more curvature. When we count the number of extrema in our bisectors, we find that while the median and modal number of extrema is 1, 11 stars have bisectors with two or more extrema. To more easily compare between stars, we again show the bisectors in the right panels of Figure \ref{fig:bisectors_Next}, sorted by the number of extrema such that bisectors with no extrema are on the top left, and bisectors with three or more extrema are in the bottom right. For clarity, the number of extrema is shown in a text box in each panel. In these panels we present depth within the profile as $(F-F_{\rm min})/(1-F_{\rm min})$ where $F$ is the flux of the line profile, and $F_{\rm min}$ is the minimum flux. This ensures that the continuum is at 1.0, and the bottom of the line is at 0.0. Additionally, because of range of bisector values is different from star to star, we normalize the velocity scale by the maximum absolute deviation from 0, $|v|_{\rm max}$. 

The bisectors with one or fewer extrema appear qualitatively similar in shape to the ``classic'' supergiant bisectors. However, we find no trend in how the bisectors are shaped (forward versus reverse C) with effective temperature. This is also true of the bisectors with more than one extremum. We searched for trends between our other observed quantities and the number of extrema, the values of these extrema, the difference between the highest and lowest extrema, and the depths where they occur, but did not find any clear trends. 

We cannot conclusively say much about the physical origin of the bisectors at this time. One possible explanation for the diversity of bisector shapes could be pulsations; YSGs have high $L/M$ ratios, and are expected to exhibit pulsations on a variety of timescales \citep{glatzel24}, and pulsations have long been put forward as a potential mechanism behind macroturbulent broadening. Time series spectra that we recently obtained of a subset of this sample will reveal whether or not we can detect (potentially periodic) variability in the bisector shapes (Dorn-Wallenstein et al. {\it in prep}). Another explanation proposed by \citet{gray86}, at least for the leftward-leaning bisectors, could be mass loss. Very little effort has been put to studying YSG winds except in the most extreme stars (hypergiants like $\rho$ Cas or IRC+10420), or stars with obvious mid-infrared excesses \citep[e.g.][who find mass loss rates of order $10^{-6}$ to $10^{-5}$ $M_\odot$ yr$^{-1}$]{humphreys23}. Modern mass loss prescriptions for stellar evolution codes simply interpolate between hot star and cool star wind models in the YSG regime \citep[e.g.][]{ekstrom12}. Depending on the evolutionary state of a given star, and the optical thickness of its wind, strong mass loss could contribute to the leftward-leaning bisectors. Finally, the anomalous bisectors may originate due to our upstream analysis. Our line profiles are derived using individual spectral lines formed at a variety of photospheric depths. As the bisector is a probe of depth-dependent velocity fluctuations, it is possible that our bisector shapes are simply due to the variable contributions of each individual line to the average profile, which we cannot disentangle. Unfortunately, a line-by-line bisector analysis is not feasible because of how broadened and therefore blended the ``individual'' spectral lines are. Ultimately, while we can only speculate, it is clear that the classical granulation scenario does not explain the array of different bisector shapes that we find.

\subsection{Turbulence and Stochastic Low Frequency Variability}\label{subsec:slfv}

\begin{deluxetable*}{llllll}
\tabletypesize{\scriptsize}
\tablecaption{Measurements of the SLF variability parameters for the stars in our sample.\label{tab:slfv_params}}
\tablehead{\colhead{Common Name} & \colhead{$\sigma_{\rm SLF}$} & \colhead{$\rho_{\rm SLF}$} & \colhead{$\tau_{\rm SLF}$} & \colhead{$\mathcal{Q}_{\rm SLF}$} & \colhead{$\log C$}\\
\colhead{} & \colhead{[x10$^{-3}$]} & \colhead{[d]} & \colhead{[d]} & \colhead{} & \colhead{}} 
\startdata
 HV 883$^\tablenotemark{a}$ & $14.36_{-0.80}^{+0.93}$ & $1.59_{-0.09}^{+0.10}$ & $0.11_{-0.01}^{+0.01}$ & $0.22_{-0.02}^{+0.02}$ & $-23.72_{-1.86}^{+2.11}$ \\ 
{[W60]} D17 & $0.25_{-0.02}^{+0.02}$ & $3.50_{-0.13}^{+0.15}$ & $2.61_{-0.45}^{+0.51}$ & $2.34_{-0.39}^{+0.44}$ & $-22.90_{-0.17}^{+0.69}$ \\ 
2MASS J05344326-6704104 & $0.60_{-0.04}^{+0.05}$ & $4.67_{-0.38}^{+0.43}$ & $1.57_{-0.35}^{+0.43}$ & $1.05_{-0.20}^{+0.25}$ & $-21.65_{-0.39}^{+0.62}$ \\ 
SP77 31-16 & $0.40_{-0.02}^{+0.02}$ & $3.86_{-0.17}^{+0.18}$ & $1.55_{-0.24}^{+0.26}$ & $1.26_{-0.18}^{+0.20}$ & $-22.75_{-0.39}^{+0.61}$ \\ 
SK -69 148 & $1.96_{-0.06}^{+0.06}$ & $1.35_{-0.04}^{+0.04}$ & $0.39_{-0.03}^{+0.03}$ & $0.90_{-0.06}^{+0.07}$ & $-22.61_{-0.48}^{+0.72}$ \\ 
RM 1-77 & $0.33_{-0.03}^{+0.03}$ & $7.70_{-0.57}^{+0.59}$ & $3.40_{-0.77}^{+0.97}$ & $1.38_{-0.30}^{+0.37}$ & $-24.47_{-1.79}^{+2.04}$ \\ 
SP77 48-6 & $0.57_{-0.03}^{+0.04}$ & $4.61_{-0.22}^{+0.24}$ & $1.84_{-0.29}^{+0.32}$ & $1.25_{-0.19}^{+0.21}$ & $-22.90_{-0.57}^{+0.82}$ \\ 
HD 269723 & $0.15_{-0.01}^{+0.01}$ & $4.14_{-0.35}^{+0.36}$ & $1.22_{-0.29}^{+0.34}$ & $0.93_{-0.20}^{+0.21}$ & $-22.65_{-0.18}^{+0.38}$ \\ 
 HV 2450 & $0.37_{-0.02}^{+0.02}$ & $3.21_{-0.18}^{+0.18}$ & $1.31_{-0.24}^{+0.26}$ & $1.28_{-0.20}^{+0.23}$ & $-22.41_{-0.41}^{+0.64}$ \\ 
HD 269879 & $0.53_{-0.04}^{+0.05}$ & $5.05_{-0.44}^{+0.46}$ & $1.96_{-0.47}^{+0.62}$ & $1.22_{-0.25}^{+0.33}$ & $-22.45_{-0.46}^{+0.70}$ \\ 
HD 269110 & $0.46_{-0.01}^{+0.01}$ & $2.12_{-0.05}^{+0.05}$ & $1.11_{-0.11}^{+0.11}$ & $1.64_{-0.17}^{+0.17}$ & $-25.17_{-1.69}^{+1.63}$ \\ 
HD 269070 & $0.50_{-0.02}^{+0.02}$ & $1.15_{-0.02}^{+0.02}$ & $1.16_{-0.16}^{+0.19}$ & $3.16_{-0.46}^{+0.55}$ & $-22.47_{-0.55}^{+0.80}$ \\ 
HD 268828 & $0.18_{-0.02}^{+0.04}$ & $1.30_{-0.37}^{+0.00}$ & $79.35_{-79.26}^{+64.74}$ & $191.68_{-191.38}^{+156.56}$ & $-24.06_{-1.74}^{+1.72}$ \\ 
HD 268865 & $0.08_{-0.07}^{+0.08}$ & $4.04_{-3.46}^{+336.60}$ & $0.05_{-0.05}^{+5.36}$ & $0.07_{-0.07}^{+1.60}$ & $-22.03_{-0.54}^{+0.80}$ \\ 
HD 269953 & $0.34_{-0.01}^{+0.01}$ & $0.76_{-0.02}^{+0.02}$ & $0.18_{-0.02}^{+0.02}$ & $0.72_{-0.08}^{+0.08}$ & $-23.19_{-0.21}^{+0.36}$ \\ 
CPD-69 496 & $0.01_{-0.01}^{+0.15}$ & $2.70_{-2.24}^{+1.34}$ & $0.15_{-0.15}^{+60.65}$ & $1.00_{-0.99}^{+77.03}$ & $-22.42_{-1.22}^{+1.12}$ \\ 
HD 268819 & $0.11_{-0.01}^{+0.01}$ & $0.78_{-0.29}^{+3.59}$ & $0.02_{-0.01}^{+0.98}$ & $0.06_{-0.02}^{+0.58}$ & $-23.23_{-0.50}^{+0.70}$ \\ 
HD 268687 & $2.19_{-0.06}^{+0.06}$ & $2.15_{-0.06}^{+0.06}$ & $0.49_{-0.03}^{+0.04}$ & $0.72_{-0.04}^{+0.05}$ & $-21.92_{-0.30}^{+0.50}$ \\ 
HD 270050 & $0.00_{-0.00}^{+0.01}$ & $0.77_{-0.45}^{+3.53}$ & $0.02_{-0.01}^{+0.03}$ & $0.10_{-0.09}^{+0.13}$ & $-22.29_{-0.51}^{+0.74}$ \\ 
HD 269697 & $0.01_{-0.01}^{+0.02}$ & $0.55_{-0.24}^{+0.34}$ & $0.01_{-0.01}^{+0.04}$ & $0.05_{-0.04}^{+0.16}$ & $-22.71_{-0.38}^{+0.56}$ \\ 
HD 269902 & $1.35_{-0.06}^{+0.07}$ & $3.47_{-0.21}^{+0.23}$ & $0.32_{-0.03}^{+0.04}$ & $0.29_{-0.03}^{+0.03}$ & $-24.82_{-1.17}^{+1.44}$ \\ 
HD 269331 & $0.73_{-0.03}^{+0.03}$ & $5.29_{-0.19}^{+0.20}$ & $2.08_{-0.23}^{+0.25}$ & $1.23_{-0.13}^{+0.14}$ & $-22.54_{-0.34}^{+0.56}$ \\ 
HD 269840 & $0.37_{-0.01}^{+0.01}$ & $1.27_{-0.06}^{+0.06}$ & $0.46_{-0.06}^{+0.07}$ & $1.13_{-0.12}^{+0.15}$ & $-22.47_{-0.38}^{+0.57}$ \\ 
CD-69 310 & $1.10_{-0.04}^{+0.04}$ & $1.37_{-0.11}^{+0.13}$ & $0.13_{-0.02}^{+0.02}$ & $0.29_{-0.03}^{+0.03}$ & $-21.29_{-0.24}^{+0.42}$ \\ 
HD 269982 & $0.48_{-0.04}^{+0.04}$ & $5.83_{-0.49}^{+0.57}$ & $1.67_{-0.35}^{+0.43}$ & $0.90_{-0.17}^{+0.21}$ & $-22.63_{-0.71}^{+1.00}$ \\ 
HD 269857 & $0.51_{-0.03}^{+0.04}$ & $4.10_{-0.39}^{+0.43}$ & $0.95_{-0.22}^{+0.27}$ & $0.73_{-0.13}^{+0.16}$ & $-21.87_{-0.31}^{+0.52}$ \\ 
HD 269651 & $0.82_{-0.04}^{+0.05}$ & $4.44_{-0.24}^{+0.25}$ & $1.42_{-0.23}^{+0.27}$ & $1.00_{-0.15}^{+0.17}$ & $-22.79_{-0.81}^{+1.09}$ \\ 
HD 269662 & $4.09_{-0.20}^{+0.22}$ & $3.72_{-0.17}^{+0.18}$ & $0.68_{-0.07}^{+0.08}$ & $0.58_{-0.06}^{+0.06}$ & $-20.81_{-0.24}^{+0.42}$ \\ 
HD 33579 & $0.80_{-0.04}^{+0.05}$ & $3.59_{-0.16}^{+0.17}$ & $0.98_{-0.14}^{+0.16}$ & $0.85_{-0.11}^{+0.13}$ & $-22.98_{-0.27}^{+0.46}$ \\ 
HD 269807 & $1.01_{-0.04}^{+0.04}$ & $1.43_{-0.05}^{+0.05}$ & $0.47_{-0.05}^{+0.05}$ & $1.03_{-0.09}^{+0.10}$ & $-22.22_{-0.66}^{+0.94}$ \\ 
HD 269604 & $1.02_{-0.06}^{+0.07}$ & $5.32_{-0.24}^{+0.25}$ & $2.46_{-0.40}^{+0.52}$ & $1.45_{-0.22}^{+0.28}$ & $-23.55_{-1.23}^{+1.49}$ \\ 
HD 269762 & $4.08_{-0.21}^{+0.23}$ & $4.36_{-0.19}^{+0.20}$ & $1.27_{-0.17}^{+0.21}$ & $0.92_{-0.12}^{+0.14}$ & $-20.84_{-0.42}^{+0.66}$ \\ 
\enddata
\tablenotetext{a}{HV 883 is a Cepheid variable with a 134-day period \citep{shapley51}. While high-amplitude, long timescale periodicities are not captured by the \tess~2-minute data, the SPOC detrending of the lightcurve of HV 883 may have imparted systematics at shorter timescales, resulting in the anomalously high value of $\sigma_{\rm SLF}$}
\end{deluxetable*}

Stochastic low frequency (SLF) variability in massive stars was discovered by \citet{blomme11}, using a small sample of O stars observed by {\it CoRoT}. \citet{bowman19b} demonstrated that SLF variability can be found in the lightcurves of all OB stars, and \citet{dornwallenstein20b} showed that this trend extended across the HR diagram to yellow and red supergiants. While the origin of the SLF variability in cool supergiants is unclear, multiple theories for the physical mechanism driving this variability in OB stars have been put forward, including internal gravity waves (\citealt{bowman19,bowman20}, but see also \citealt{anders23}), subsurface convection (\citealt{cantiello21,schultz22} but see also \citealt{bowman24}), and wind-driven processes \citep{krticka21}. These mechanisms have also been invoked to explain macroturbulence in OB stars, implying a connection between the two phenomena. Indeed, \citet{bowman20} demonstrated that there is a clear link between the two processes: $\vturb$ is positively correlated with the amplitude of the SLF variability, and negatively correlated with the characteristic frequency (i.e., the location of the break in the ``red noise'' power spectrum of the variability). Apparently, the mechanism that is driving the SLF variability is also responsible for creating the large scale tangential motions associated with macroturbulence. This is perhaps unsurprising, given that macroturbulence is associated with large-scale inhomogeneities in the photospheric velocity field that could result in photometric variability.

Because our sample of stars were all studied by \citetalias{dornwallenstein22}, by construction all of them were observed at two-minute cadence by \tess. In order to use the most updated data, we accessed all available target lists for \tess as of 13 December, 2024,\footnote{\url{https://tess.mit.edu/public/target_lists/target_lists.html}} and downloaded all two-minute cadence data for our stars. In practice, because our stars are all in the \tess~southern continuous viewing zone, this amounts to data from \tess~sectors 1-13 and 27-39 (the sectors analyzed by \citetalias{dornwallenstein22}), as well as sectors 61-69. All the {\it TESS} data used in this paper can be found in the Barbara A. Mikulski Archive for Space Telescopes (MAST): \dataset[10.17909/t9-nmc8-f686]{http://dx.doi.org/10.17909/t9-nmc8-f686}. These data were processed identically to \citetalias{dornwallenstein22}: we used the {\tt PDCSAP\_FLUX} lightcurves, and stitched together data from different \tess~sectors by dividing each sector’s data by the median flux in each sector. Finally, to increase our signal-to-noise, we apply a rolling 30-minute Gaussian smooth to the data. This allows us to only use every 15$^{\rm th}$ measurement (i.e., one point every 30 minutes) in the lightcurve, as the inference procedure described below is computationally expensive and the smoothed data are artificially correlated on shorter timescales.

All of the lightcurves in our sample show SLF variability (i.e., have rising power at low frequencies). Multiple methods exist to characterize SLF variability, all broadly focused on measuring characteristic timescales and amplitudes of the variability. For example, \citet{dornwallenstein20b} fit Lomb-Scargle periodograms \citep{lomb76,scargle82} with a quasi-Lorentzian function, which is also used to characterize granulation in sunlike stars \citep[e.g.][]{harvey85,kallinger14}. However, \citet{bowman22} demonstrated that this approach only yields reliable results when the amplitude of the SLF variability is much larger than the noise floor of the data. Instead, they suggest a Gaussian process \citep[GP;][]{rasmussen06} based approach, using the Python package {\sc celerite2} \citep{foremanmackey17,foremanmackey18} with a {\tt SHOTerm} kernel (see \citealt{bowman22} and the {\sc celerite2} documentation for more details). A notable feature of this model is that it allows one to infer an amplitude, $\sigma_{\rm SLF}$, a characteristic timescale corresponding to the quasi-period of the perturbations, $\rho_{\rm SLF}$, and a timescale on which perturbations are damped, $\tau_{\rm SLF}$. The latter two quantities can be combined to define the quality factor, $\mathcal{Q_{\rm SLF}}=\pi\tau_{\rm SLF}/\rho_{\rm SLF}$, which can be used to indicate how stochastic the variability is. Variability with $\mathcal{Q}_{\rm SLF}<1/2$ is strongly damped, variability with $\mathcal{Q}_{\rm SLF}>1/2$ is underdamped, and variability $\mathcal{Q}_{\rm SLF}\gg 1$ is highly periodic.

After subtracting off the periodic signals recovered from the data by \citet{dornwallenstein22}, we used the {\sc Jax} \citep{jax18} backend of {\sc celerite2}, along with {\sc NumPyro} \citep{phan19} to perform Hamiltonian Monte Carlo \citep{duane87} using No U-Turn Sampling (NUTS, \citealt{hoffman14}), in order to measure the SLF hyperparameters for each star. We adopt wide uniform priors in log-space for each parameter, and include a ``jitter'' parameter ($C$) to account for excess white noise in the data. We use 4 chains and perform 2000 ``warm-up'' steps where the NUTS kernel learns the optimal step size and mass matrix for the sampling, and identifies the neighborhood of parameter space containing the {\it a posteriori} parameter values. After warm-up, we take 10,000 samples per chain. Table \ref{tab:slfv_params} lists the inferred values and errors for $\rho_{\rm SLF}$, $\tau_{\rm SLF}$, and $\sigma_{\rm SLF}$, as well as $\mathcal{Q}_{\rm SLF}$ and $\log C$, computed from the median posterior sample and the 16$^{th}$ and 84$^{th}$ percentiles.

We then searched for any correlations between the SLF hyperparameters and the macroturbulent velocity, again using a Spearman rank correlation test. After an exhaustive search, we found statistically significant correlations between $\vturb$ and both $\sigma_{\rm SLF}$ ($R=0.496$, $p=0.031$) and $\mathcal{Q}_{\rm SLF}$ ($R=-0.461$, $p=0.047$) that indicate that in stars with more macroturbulence, the SLF variability is higher amplitude, and more stochastic. No difference was found between groups A and B. We also searched for correlations between the SLF hyperparameters and the line profile asymmetry as measured by both $|R_{sk}-0.45|$ and $|\sqrt[3]{\vthird}|$ as another way of probing coherent movement on large spatial scales. Here, the behavior diverges between the two groups of YSGs. In group A, we found a statistically significant negative correlation between relative skewness and $\sigma_{\rm SLF}$ ($R=-0.650$, $p=0.022$). In group B we found a statistically significant positive correlation between relative skewness and both $\tau_{\rm SLF}$ ($R=0.857$, $p=0.014$) and $\mathcal{Q}_{\rm SLF}$ ($R=0.857$, $p=0.014$), and a statistically significant negative correlation between $|\sqrt[3]{\vthird}|$ and $\mathcal{Q}_{\rm SLF}$ ($R=-0.857$, $p=0.014$).

\section{Discussion}\label{sec:discuss}

\subsection{Two Regimes of Turbulence --- Comparison with Known Post Red Supergiants \& Binaries}

One explanation for the existence of two groups of YSGs in $\vturb$-$\vmicro$ space is that one of the groups consists of ``normal'' YSGs crossing the HR diagram for the first time, and the other consists of YSGs in a different evolutionary state. As this survey was started with the goal of identifying post-RSGs, we first compare our sample with a sample of known post-RSGs. Post-RSGs have historically been identified via the presence of circumstellar material \citep[e.g.][]{jones93,humphreys97,humphreys02,shenoy16}, drastic photometric variability induced by dynamical instabilities \citep[e.g.][]{nieuwenhuijzen95,stothers01}, or spectroscopic signatures \citep[e.g.][]{humphreys13,gordon16,kourniotis22}. While less-luminous post-RSGs do not necessarily display these ``smoking gun'' signatures of their evolutionary status (a fact that inspired this survey), we can still turn to the small overlap between our sample and vetted samples of post-RSGs.

\citet{humphreys23} conducted a survey of the most luminous LMC YSGs, and identified 9 candidate post-RSGs, of which 6 are included in our survey: HD 268687, HD 269953, SK -69 148, CD-69 310, HD 269723, and HD 269840. Among this group of stars, only HD 269953 belongs to group B. The rest of the candidate post-RSGs are in group A. Of the candidate post-RSGs in our sample, three stars were specifically identified by \citeauthor{humphreys23} as having substantial circumstellar dust; HD 269953 (group B), and SK -69 148 and HD 269723 (group A). 

It is important to note that one of the criteria that \citeauthor{humphreys23} used to identify candidate post-RSGs is that they were identified by \citet{dornwallenstein22} as being fast yellow pulsating supergiants (FYPS). While \citet{dornwallenstein22} did identify a statistically significant overdensity of pulsators in the upper HR diagram that they associated with candidate post-RSGs, they, as well as \citet{pedersen23}, identified issues with contamination in the \tess~lightcurves that renders the ``FYPS'' label useless for confirming the evolutionary status of any individual star until it can be confirmed by additional follow-up observations. Regardless, two of the three ``smoking gun'' candidate post-RSGs from \citet{humphreys23} belong to group A, and the other star (HD 269953) is both very luminous and appears to be in an interacting binary system \citep{kourniotis22}. Of course, these are small number statistics, so little in the way of concrete conclusions can be made from these data alone. We also compared our sample with the catalog of YSG+OB binaries from \citet{ogrady24}, though the overlap between our sample and the one from \citeauthor[]{ogrady24} is small. Only one star was identified by \citeauthor{ogrady24} as being a candidate binary: HD 268828, which belongs to group A.

\subsection{Turbulence in Model YSGs}

\begin{figure*}[t!]
\centering
\includegraphics[width=0.9\textwidth]{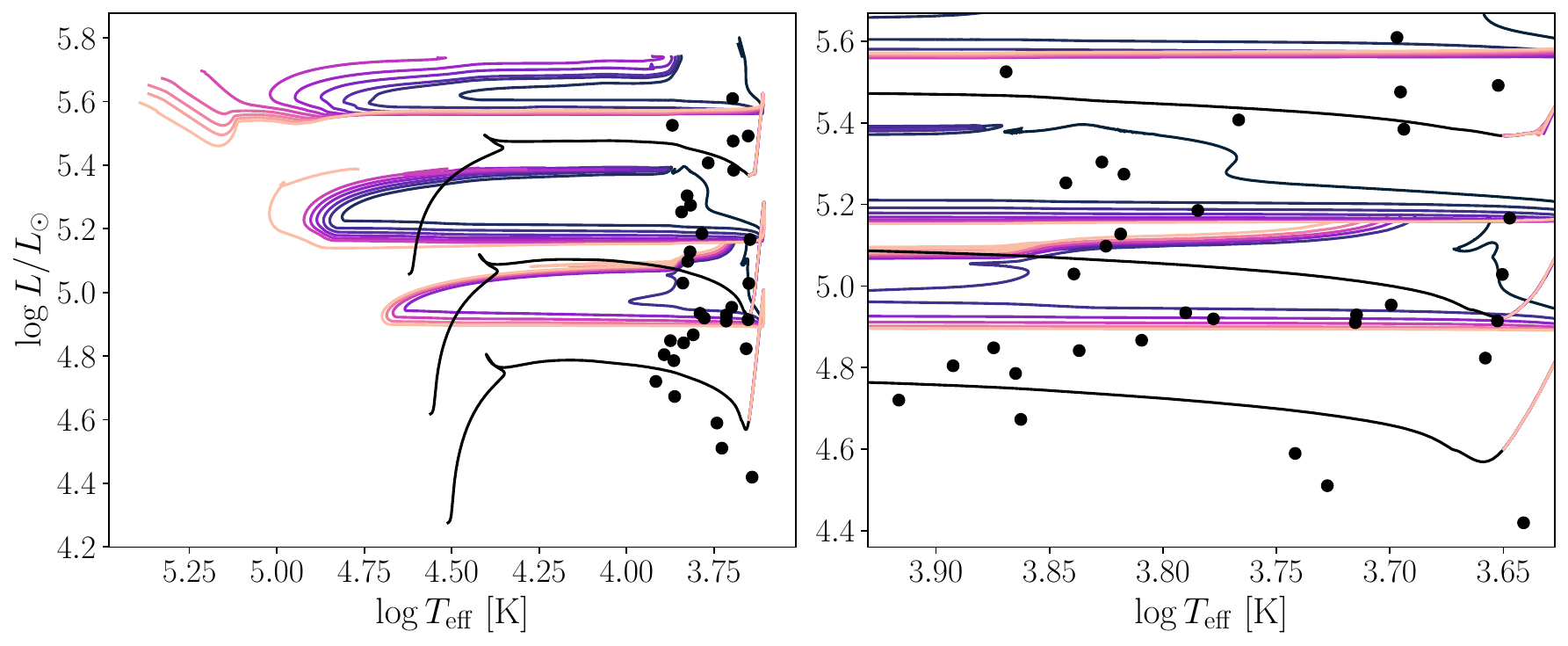}
\caption{Hertzprung-Russell diagram showing the evolution of our $\alpha_{\rm MLT}=2.1$ {\sc MESA} models. The pre-RSG phase is colored in black, after which tracks are colored by the degree of stripping that occurs during the RSG phase, with stripping increasing from dark to light pink.}\label{fig:mesa_HR}
\end{figure*}

We now wish to understand our measurements in the context of stellar structure and evolution. In particular, the models we used to fit our profiles in \S\ref{subsec:profilefit} are phenomenological models developed under idealized assumptions about turbulent motions on various length scales. The physical origins of the line broadening processes that we have characterized as ``macroturbulence'' and ``microturbulence'' are still unclear. The first place to look (and indeed, the physical process that is most explored in the literature as an origin for these phenomena) is convection. Convection is a turbulent process that occurs across a wide range of size scales, and is well-studied using a variety of theoretical models and tools. Therefore, we chose to make the convective properties of stellar structure models our first point of comparison with our $\vturb$ and $\vmicro$ measurements.

We use {\sc MESA} \citep{paxton11,paxton13,paxton15,paxton18,paxton19,jermyn2023} version r23.05.1 to produce 1D stellar structure models for YSGs, focusing on the stripped-envelope case. We study single stars with initial masses of $M_i = 15$, $20$, and $30\, M_{\odot}$ as they cross the Hertszprung gap during the first YSG phase just after leaving the main sequence. 
We include time-dependent convection (TDC; \citealt{kuhfuss86}), which reduces to Mixing Length Theory (MLT; \citealt{bohmvitense58,cox68}, see recent review by \citealt{joyce23}) over long timescales \citep{jermyn2023}. 
Because we are interested in the behavior of the convection in our models --- which is particularly sensitive to the treatment of mixing --- we compute models for three values of the mixing length parameter, $\alpha_{\rm MLT} = 1.5$, 2.1, and 2.7. We take $\alpha_\mathrm{MLT}=2.1$ as fiducial to approximately fit the LMC following \citet{chun18}. All models are initialized with a metallicity of $Z=0.006$, suitable for the LMC \citep{eggenberger21}. 
Inlists follow closely the setup from Section 7.2 of \citep{jermyn2023}, using the \citet{brott11b} wind prescription, with step overshooting over the H core and exponential overshooting over all shells. This treatment tends to lead to larger cores and few blue loops. The same physics are used across all models.\footnote{We make our {\sc MESA} inlists containing all of the included physics --- including rotation, mixing (including overshooting), and winds --- available on Zenodo at 10.5281/zenodo.15786147}

From this set of models, we then produce stellar models of stripped post-RSGs by artificially removing mass via strong winds of $\dot{M}\gtrsim 10^{-6}$--$10^{-5}\, M_{\odot}/\mathrm{yr}$ from the red supergiant phase onward, with a small number of timesteps during the RSG phase reaching mass loss rates as high as $\sim10^{-4}\, M_{\odot}/\mathrm{yr}$. This ``stellar engineering'' method can also mimic case C mass transfer (i.e., mass transfer after the onset of core He burning; \citealt{lauterborn70}), at least from a stellar structure standpoint. More specifically, these winds are initiated after the start of core He burning. This process is repeated for varying magnitudes of the imposed wind mass loss, resulting in stripped post-RSGs which pass through a second YSG phase during the stripping process. During this stage, the envelopes reach a range of $^1$H masses between $\approx 0.05$--$0.7\, M_{\odot}$, from initial envelope masses of $6$--$9\, M_{\odot}$. Mass loss is implemented as a scaling factor applied to the wind mass loss scheme adopted from \citet{brott11a,brott11b}. While we use a variety of scaling factors between 1 and 4, the range of scaling factors required to strip sufficient envelope for the star to appear yellow is generally higher in the lower-mass models. This is not surprising because in the underlying mass loss scheme, $\dot{M}\propto L^\alpha$ for some positive power $\alpha$, i.e., the more massive and luminous models lose more mass. Therefore, to achieve qualitatively similar degrees of envelope stripping in two models with different initial masses, a larger scale factor is needed in the lower-mass model.  

Combined, these models span the temperature, luminosity, and initial mass range of our sample, and our variations in input physics (in particular, $\alpha_{\rm MLT}$ and the strength of the imposed mass loss) entail a range of modeling uncertainties, especially in the characteristic velocity of convection which is directly a function of the adopted mixing length and envelope mass. We take these models to be illustrative, rather than prescriptive, and make comparisons to the structure (as a function of envelope mass scenario) rather than the evolutionary pathways. 

As discussed in \S\ref{sec:intro}, RSGs may undergo similar envelope stripping through some mechanism that induces strong mass loss, such as binary interactions. Our mass loss implementation is agnostic to the exact physical process, so long as it occurs after the He core is fully-formed. As the first YSG phase occurs before the onset of core He burning, each stripped RSG experiences the same pre-RSG evolution given the same initial mass. In the $\alpha_{\rm MLT}=2.1$ models, for stars with $M_i = 15/20/30\, M_{\odot}$, the first YSG phase duration is $\sim 4400/3100/2300$ yr, respectively, roughly the Kelvin-Helmholtz time of the contracting He core after the cessation of core H-burning. Given the consistent physical origin and insensitivity of the core structure to the convective efficiency \citep{joyce23}, these lifetimes are consistent across varying values of $\alpha_{\rm MLT}$. We also note that, at larger $\alpha_{\rm MLT}$ values, the radius of a fully convective envelope decreases. This causes the Hayashi limit to shift to warmer temperatures that intersect the temperature range of our sample, which may increase the calculated YSG lifetime in higher-$\alpha$ models. In general, the second YSG phase of the stripped post-RSGs lasts $\sim 2$--$8\times 10^4$ yr. We verify that the stripped models presented here are in approximate hydrostatic and thermal equilibrium.

Figure \ref{fig:mesa_HR} shows the evolution of our $\alpha_{\rm MLT}=2.1$ models in the Hertzprung-Russell diagram. The pre-RSG phase is colored in black for all three initial masses. From there, tracks are colored by the mass loss scaling factor. To capture the fact that different initial masses used different ranges of mass loss scaling factors, we separately color the tracks for each initial mass to show the range of wind scaling factors used for that initial mass only. This means that, for each initial mass, the lightest pink track has the most imposed mass loss, while the darkest purple track has the least, and similarly colored tracks across different initial masses have qualitatively similar degrees of envelope stripping. The locations of the YSGs in our sample are shown as black points. The left side of the figure shows the complete evolutionary tracks, while the right size is zoomed in to the subspace containing our sample.

All of the model YSGs, both stripped and un-stripped, possess convective envelopes while in the region of the HR diagram containing our sample. As convection is a turbulent process, we wish to compare the properties of the convection in the models with our observed turbulent velocities. To estimate an observable turbulent velocity from our models, we first define the convective velocity:
\begin{equation}
    v_{\rm conv} = \bigg(\frac{L_{\rm conv}}{4\pi\rho r^2}\bigg)^{1/3}
\end{equation}
where $L_{\rm conv}$ is the luminosity carried by convection at radius $r$. We then compute the mass-averaged convective velocity within the convective zone, $\langle v_{\rm conv}\rangle$ (e.g., \citealt{wu21}). 

\begin{figure*}[t!]
\centering
\includegraphics[width=0.9\textwidth]{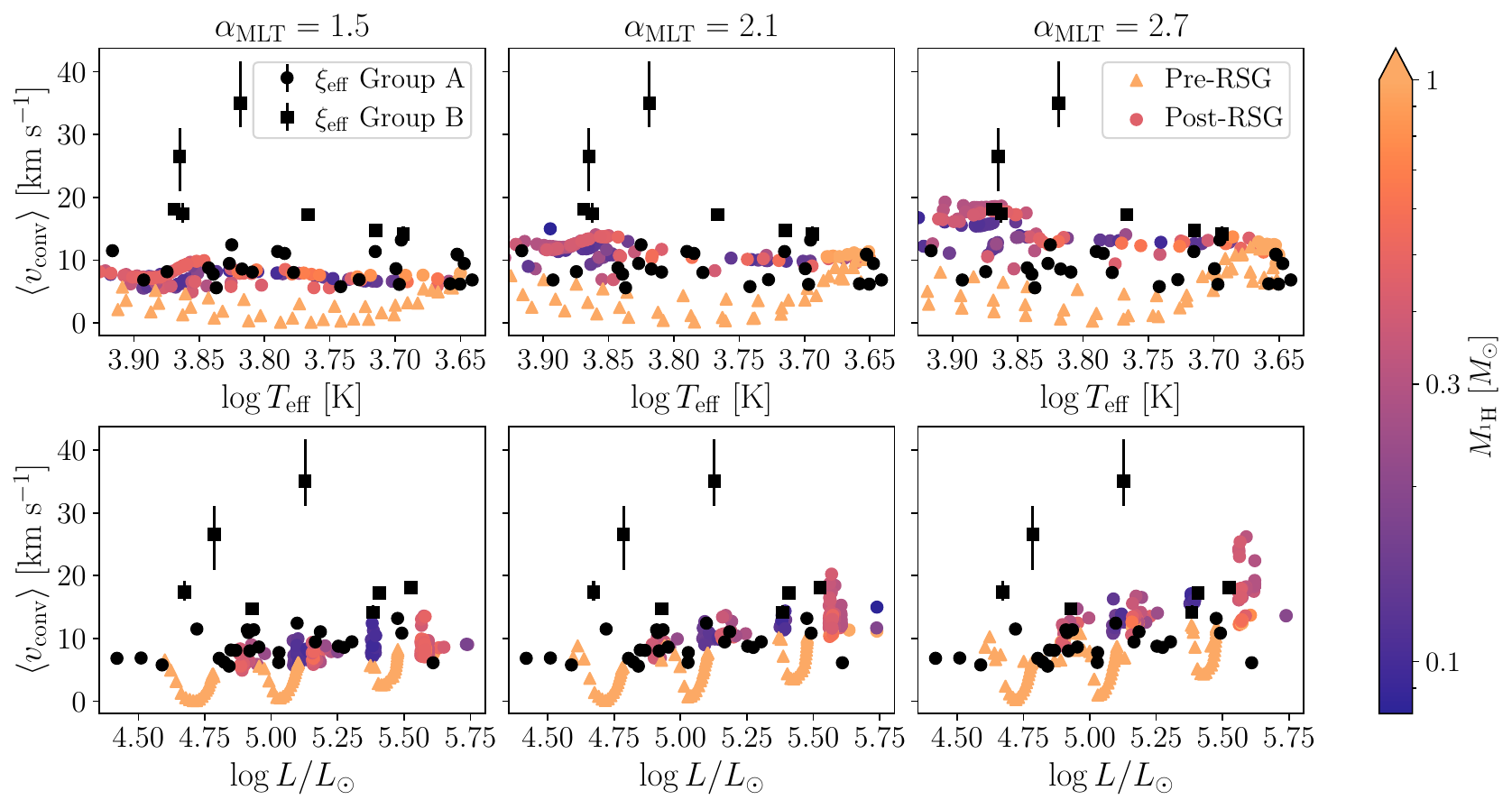}
\caption{Comparison between the average convective velocity in our {\sc MESA} models and the observed microturbulence, $\vmicro$, black points, as a function of $\teff$ (top row) and $\lum$ (bottom row), for $\alpha_{\rm MLT}=1.5$ (left), 2.1 (center), and 2.7 (right). Model points are colored by their remaining $^1$H envelope mass as indicated by the colorbar.}\label{fig:avgvconv_vmic}
\end{figure*}

Generally, the model $\langle v_{\rm conv}\rangle$ values are $\sim$10 km s$^{-1}$, a factor of $\sim$5 below our measured $\vturb$ values, but consistent with our $\vmicro$ measurements. Figure \ref{fig:avgvconv_vmic} shows the average convective velocity in the envelope of each YSG model, compared to the derived microturbulent velocities from the sample stars, as a function of both $\teff$ (top row), and $\lum$ (bottom row), for three values of $\alpha_{\rm MLT}$ as shown at the top of each column. Model $\langle v_{\rm conv}\rangle$ values are colored by the amount of $^1$H remaining in the envelope, such that the pre-RSG models are orange, and darker coloring indicates increased stripping. To further distinguish between the pre- and post-RSG timesteps, pre-RSG timesteps are shown as triangles, and post-RSG timesteps as circles. Observations of Group A YSGs are shown as black circles, while group B YSGs are black squares.

In Figure \ref{fig:avgvconv_vmic}, we find good agreement between $\langle v_{\rm conv}\rangle$ in our models and our observed $\vmicro$ values for group A YSGs. This is true both for the overall values as well as how they trend with observed stellar properties. Furthermore, there is a separation between the convective velocities of $\lesssim 10$ km/s exhibited during the first, pre-RSG YSG phase and the $10$--$20$ km/s convective velocities achieved during the second, post-RSG YSG phase. This is due to a few factors. First, the model YSGs have higher luminosities during the post-RSG phase, so the combination of convection and radiative diffusion together must carry a larger flux. 
Moreover, in the stripped case, there is less mass in the envelope (by 1-2 orders of magnitude), so at fixed luminosity and temperature, the convective velocity must be faster in order to transport comparable convective flux (as $F_{\rm conv}\sim \rho v_{\rm conv}^3$). 
There is some nuance here, because as the optical depth decreases, radiation begins to carry more flux and 3D effects become significant for stars near the Eddington limit (see, e.g. discussions by \citealt{jiang15,schultz20,goldberg22}). 
But as long as convection is carrying significant flux, then the convective velocities are accordingly larger in the stripped case. 

Notably, the $\vmicro$ values of the group B YSGs are higher than the mass-averaged convective velocities of even the most stripped models by a fair amount. 
The trends in the properties of convection persist across the different $\alpha_{\rm MLT}$ values, though the overall normalization is slightly higher/lower for $\alpha_{\rm MLT}=2.7/1.5$ respectively. This difference is a direct result of mixing length theory: $v_{\rm conv}\propto F_{\rm conv}^{1/3}$, and $F_{\rm conv}\propto\alpha_{\rm MLT}$, i.e., $v_{\rm conv}$ has a weak positive dependence on $\alpha_{\rm MLT}$. 

We pause also to note that convection is an inherently three-dimensional process, wheres the models presented here are fundamentally 1D. The velocity ``predicted'' by mixing length theory is really a \textit{characteristic} velocity of convective transport, whereas the physical velocities will span a range of values \citep[e.g.][]{stein89,stein98}. 
Moreover, MLT is fundamentally a local theory, i.e. the velocities are calculated from fluxes at each given radius, whereas 3D models of luminous cool supergiants exhibit large-scale global flows \citep[e.g.][]{brun09,chiavassa11,chiavassa24,goldberg22,antoni22,ma24,ma25} since the pressure scale height and thereby mixing length is a significant fraction of the envelope. 

\begin{figure*}[t!]
\centering
\includegraphics[width=0.9\textwidth]{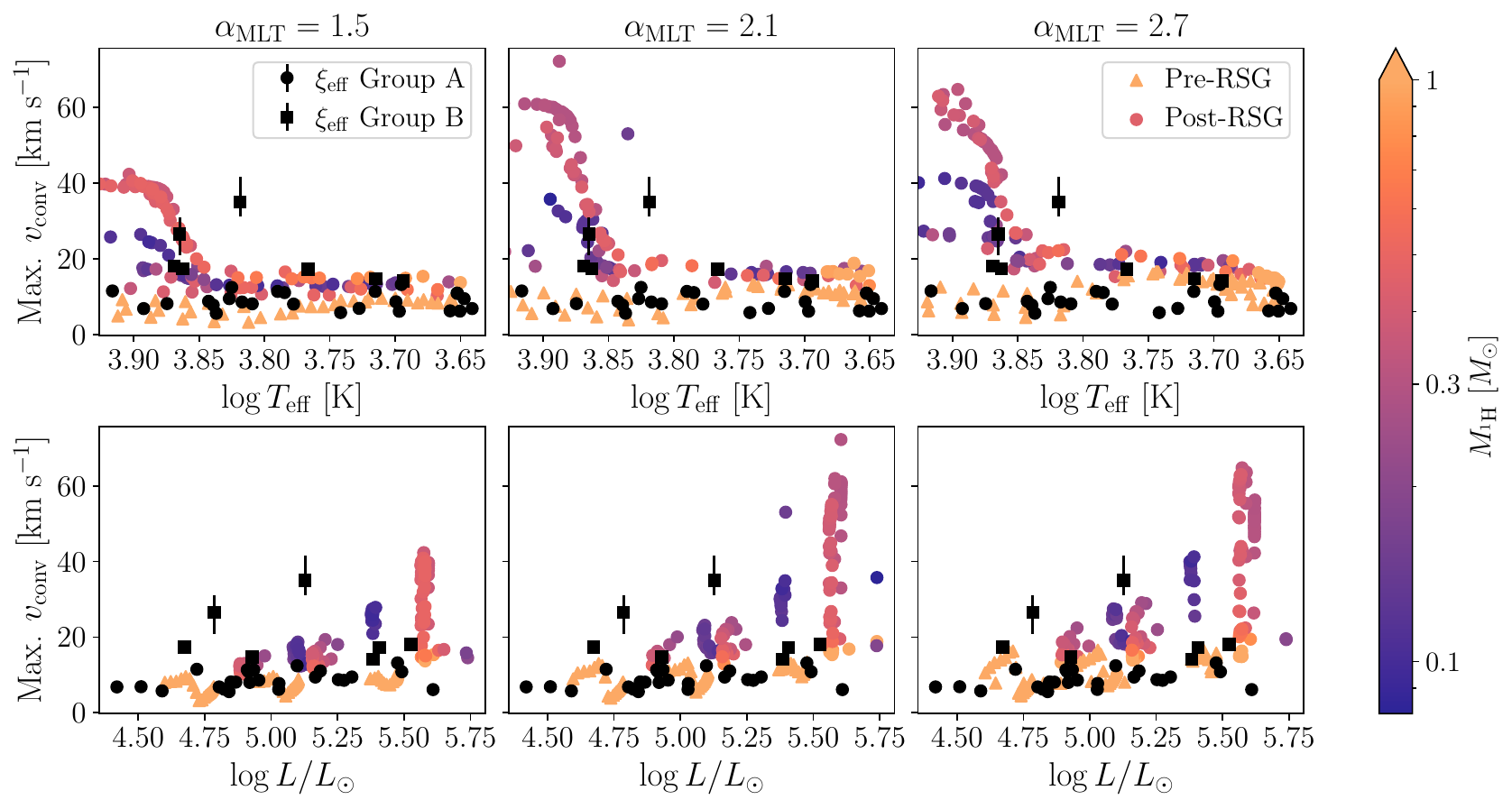}
\caption{Similar to Figure \ref{fig:avgvconv_vmic}, but showing the maximum convective velocity in the envelopes of our models.}\label{fig:maxvconv_vmic}
\end{figure*}

Given that the observed flows could be sourced from global properties of the stellar envelopes, we also compare to the maximum convective velocity anywhere in the envelope. 
Figure \ref{fig:maxvconv_vmic} shows a comparison between our observed $\vmicro$ and the maximum $v_{\rm conv}$ value in the envelopes of our models. We find a much larger difference between the behavior of the pre- and post-RSG YSG models, especially at temperatures above $\teff\approx3.8$. Furthermore, the maximum convective velocities in the pre-/post-RSG models agree favorably with the $\vmicro$ values observed in group A/B respectively. This is true both for the magnitude of the observations, and how they trend with stellar properties, particularly the sharp increase in $\vmicro$ around $\teff\approx3.85$ in the group B YSGs. We pause to note that this temperature is also where the yellow void occurs in the HR diagram. Furthermore, the bottom panels of Figure \ref{fig:maxvconv_vmic} show that the sharp increase in the maximum $v_{\rm conv}$ values becomes increasingly more pronounced in the higher luminosity stripped YSG models.

However, we still warn that velocities from MLT should not be taken as strict predictions, though they are informative in the ways in which the different model sets differ from each other. 
In more realistic simulations of 3D convection in luminous stars \citep[e.g.][]{brun09,chiavassa11,chiavassa24,goldberg22,antoni22,ma24,ma25}, the velocities of individual parcels of moving gas can be much larger than the characteristic velocity derived from the local flux adopted by mixing length theory. 
As a result, fluid velocities from a 3D YSG atmosphere model
(e.g. \citealt[]{goldberg25b}) are expected to span values up to a few times larger than either the mass-averaged quantity shown in Figure \ref{fig:avgvconv_vmic}, or even the maximum values of $v_\mathrm{conv}$ seen in our {\sc MESA} envelope models shown in Figure \ref{fig:maxvconv_vmic}. Future calculations of line formation in these tenuous atmospheres are needed to go beyond comparisons to MLT, e.g. following the recent methods of \citet{schultz23a,ma24,delbroek25}, and others.

\begin{figure*}[t!]
\centering
\includegraphics[width=0.9\textwidth]{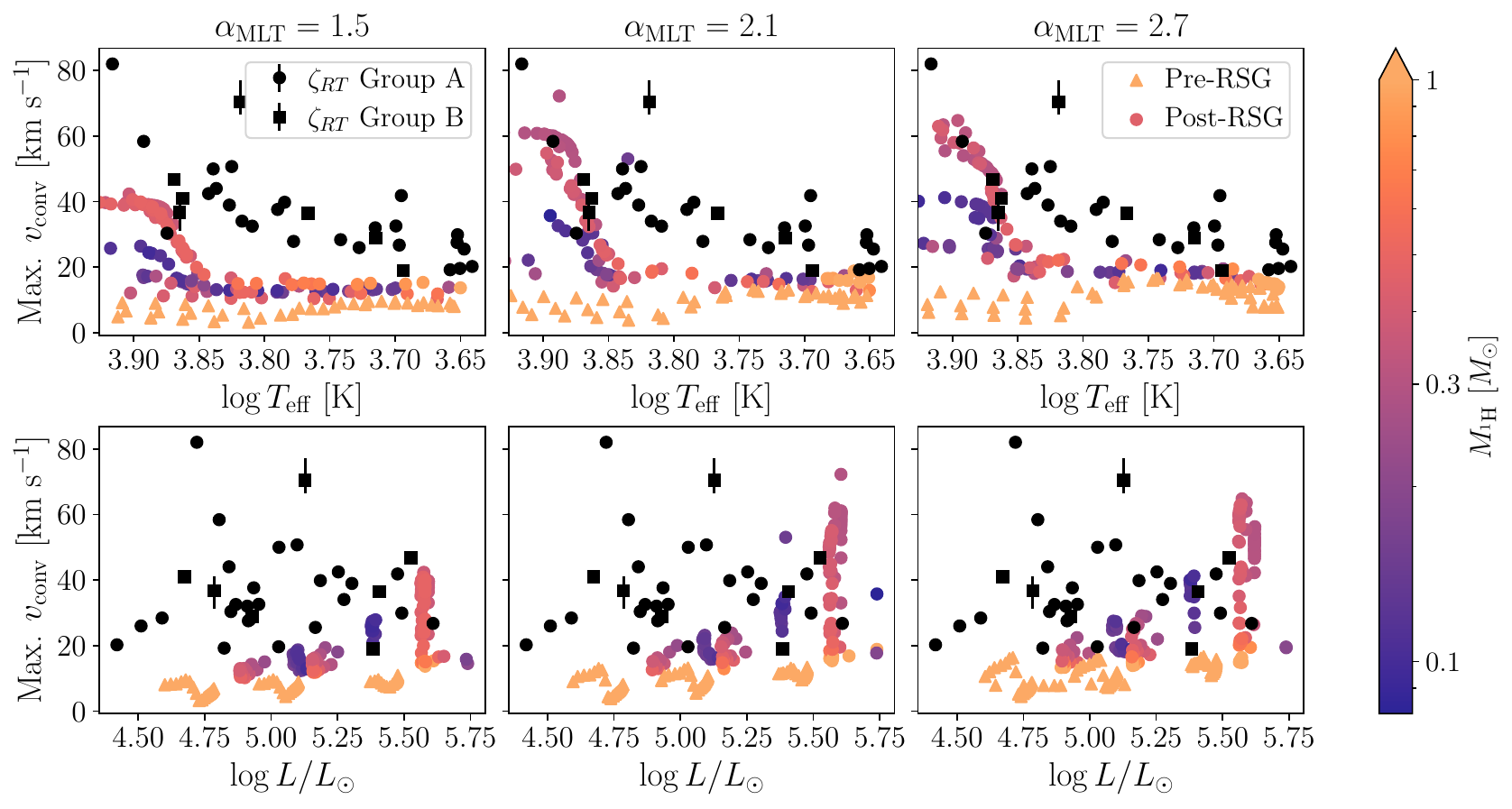}
\caption{Similar to Figure \ref{fig:maxvconv_vmic}, but showing the maximum convective velocity in the envelopes of our models compared to $\vturb$.}\label{fig:maxvconv_vRT}
\end{figure*}

While the mapping of $v_\mathrm{conv}$ as predicted by MLT to either micro- or macro-turbulent velocities is not one-to-one, 
these comparisons between the convective velocities in our 1D models and the observed microturbulent velocities are promising. On the other hand, Figure \ref{fig:maxvconv_vRT} shows a comparison between the maximum $v_{\rm conv}$ values in our models and our measured macroturbulent velocities. Our $\vturb$ measurements are much higher than the maximum convective velocities, let alone the mass-averaged quantity shown in Figure \ref{fig:avgvconv_vmic}. This is particularly true for the lower-luminosity stars in our sample. However, the rapid increase in the maximum convective velocity in the models with$\teff\gtrsim3.8$ {\it does} appear to agree with the observed trend between $\vturb$ and $\teff$.

We suggest a few possible interpretations of these comparisons:
\begin{itemize}
    \item The group A YSGs in our sample are post-RSGs. The lifetime of the post-RSG YSG phase in our models is significantly longer than the pre-RSG phase. Furthermore, the dearth of RSG progenitors of type II supernovae above $\lum\approx5$ \citep[e.g.][]{smartt15,rodriguez22}, as well as the existence of numerous YSG progenitors of type IIb supernovae with luminosities around this same threshold \citep{aldering94,crocket08,maund11,eldridge13,vandyk14} imply that strong mass loss during the RSG phase (whatever the origin between winds and binary interactions) plays a role in the evolution of a significant fraction of massive stars. If so, it may ultimately be unsurprising that, in a sample of luminous YSGs, the majority of them are partially stripped. This interpretation is supported by the agreement between $\vmicro$ and the $\langle v_{\rm conv}\rangle$ values in our post-RSG models. However, this interpretation does not explain the origin of the group B YSGs.
    \item The group B YSGs in our sample are post-RSGs, while group A YSGs are pre-RSG objects. If the stripping mechanism only occurs in a very small fraction of RSGs, then we would expect the majority of a sample of YSGs to contain objects in a pre-RSG phase, despite the prolonged lifetimes of our post-RSG YSG models. Indeed, observational surveys for post-RSG objects have only revealed small numbers of ``smoking gun'' candidate post-RSGs \citep{gordon16,humphreys23}. Furthermore, if the maximum convective velocity is a better representation of the turbulent velocity that would be observed in a real YSG, then the comparison between $\vmicro$ and the maximum $v_{\rm conv}$ values supports this interpretation. 
    \item Given the numerous uncertainties in stellar evolution (and other caveats which we detail below), there are any number of stellar engineering choices one could make to adjust the predicted velocities up or down by factors of a few. The qualitative agreement between both micro- and macroturbulent velocities and the convective velocities of our models is simply a promising indication that turbulent convection in the models may be approximately consistent with the data, and more observations are needed to understand the meaningful differences between groups A and B. Indeed, at face value, the convective velocities of our 1D models struggle to match the largest $\vturb$ values in both groups A and B, implying that further study is needed to interpret our measurements.
\end{itemize}

Of course, there are a number of important caveats to this discussion, ranging from the purely observational to the purely theoretical. First, our turbulent velocities are the measured widths of functions used to fit line profiles that are representative of the entire observed spectrum. As discussed above, this includes lines formed in a variety of depths. While the strength of the macroturbulent broadening makes an analysis of individual, unblended lines exceedingly difficult, this averaging across lines does result in a loss of information. 

Second, the energy in turbulence is distributed across a wide range of size scales \citep{kolmogorov41}, from large convective eddies down to below the photon diffusion length scale. As a result, it is somewhat simplistic to describe turbulence with two velocities ($\vmicro$ and $\vturb$) that, nominally, represent two extremely different size scales. Likewise it is simplistic to describe turbulent velocities theoretically as one $v_\mathrm{conv}$ from MLT.

Third, if our models are at all representative of realistic stars, then there should be a relationship between the observations and our predicted convective velocities --- and indeed, the agreement that we find is promising. However, given the above two caveats, it is likely that the quantities we have measured are not described exactly by the properties of our 1D models that we have extracted. 

Fourth, there exist a number of uncertainties in stellar evolution, particularly for massive stars, and especially for yellow supergiants (indeed, this is why they are a ``magnifying glass''). One main uncertainty of such stellar models is the treatment of convection via  mixing length theory. A major limitation of MLT in 1D is its inability to fully resolve behavior of the convective cells near the surface of the star. For instance, 3D simulations of turbulent convection in massive star envelopes show that convective zones may exhibit significant velocity fluctuations well past the boundary for convection defined by the default criteria of 1D stellar models \citep[e.g.][]{schultz22,schultz2023}. In the near-Eddington regime, pressures from radiation become hydrodynamically relevant to determine turbulent and wind-launching properties \citep[e.g.][]{jiang15,jiang18}. This may even entail different correlations within the near-surface convective regions: high-entropy material with low opacity can sink, dense material with higher opacity can move outward, from high optical depths out past the photosphere \citep{schultz20,schultz2023,goldberg22}. This is to say, it is all but certain that we are not modeling convection correctly. Nonetheless, we believe that we are modeling convection informatively.

Finally, the large-scale velocity fields that are responsible for broadening the observed line profiles by 50-100 km s$^{-1}$ may also come from physics that are not included in our models, including pulsations, outflows, and more. The velocity fields in a real YSG are inevitably going to be far more complex than the simple treatment we have given them in our models.

\section{Summary \& Conclusion}\label{sec:conclusion}

Our findings are as follows:
\begin{itemize}
    \item Turbulence on multiple size scales is a key source of line broadening in the spectra of LMC YSGs. For all stars, macroturbulent velocity $\vturb$ dominates over microturbulent velocity $\vmicro$. This finding neatly explains our previous finding of anomalously high surface gravities for the stars in this sample (\citetalias{chen24}). We present revised atmospheric parameter measurements in this paper.
    \item Our sample contains two groups of YSGs, defined by the relative importance of the two turbulent velocity scales.
    \item $\sim$80\% of YSGs belong to group A, in which stars show significantly more macroturbulence than microturbulence. The line profiles of these stars are also notably asymmetric.
    \item The remaining $\sim$20\% of YSGs belong to group B, in which microturbulence is increased relative to macroturbulence. The line profiles of these stars are more symmetric.
    \item We compare our results to similar studies of macroturbulence in OB stars. In many cases, the correlations we find agree with these studies (e.g., trends with temperature, trends with SLF variability amplitude). However, in a few cases --- line profile asymmetry as well as the SLF variability characteristic timescale --- the sign of these correlations differs from the OB star regime.
    \item By constructing 1D evolutionary models of YSGs in both pre- and post-RSG phases, and comparing the predicted convective velocities with our $\vturb$ and $\vmicro$ measurements, we find promising agreement between the models and $\vmicro$. This is true both in the velocity directly, and how the measurements trend with stellar properties. 
\end{itemize}

Ultimately, we find that observations of the line profiles of YSGs are a promising new avenue into understanding the structure and evolution of massive stars during this poorly-understood phase of their lives.

\newpage

\acknowledgments

The authors acknowledge that the work presented was in part conducted on occupied land originally and still inhabited and cared for by the first peoples of Los Angeles: the Tongva, Kizh, and Chumash peoples. We honor with gratitude the land itself, and the original inhabitants of these places. We are grateful to have the opportunity to live and work on their ancestral lands.

This paper includes data gathered with the 6.5 meter Magellan Telescopes located at Las Campanas Observatory, Chile. In particular, we acknowledge Las Campanas Observatories telescope operators Mauricio Martinez and Jorge Araya, and support astronomers Matias Diaz and Carlos Contreras, without whose expertise this science could not be done.

This research has made use of the SIMBAD database, operated at CDS, Strasbourg, France.

This work has made use of the VALD database, operated at Uppsala University, the Institute of Astronomy RAS in Moscow, and the University of Vienna. Ryabchikova T., Piskunov, N., Kurucz, R.L., et al., Physics Scripta, vol 90, issue 5, article id. 054005 (2015), (VALD-3).

This work made use of the following software and facilities:

\facilities{Magellan:Clay (MIKE double echelle spectrograph), \tess}. All the {\it TESS} data used in this paper can be found in MAST: \dataset[10.17909/t9-nmc8-f686]{http://dx.doi.org/10.17909/t9-nmc8-f686}, Sectors 1-13, 27-39, 61-69.

\software{
ACID v0.1.0 \citep{dolan24}
Astropy v6.1.0 \citep{astropy13,astropy18,astropy22},
Astroquery v0.4.7 \citep{astroquery},
Celerite2 v0.3.2 \citep{foremanmackey17,foremanmackey18},
Cmasher v1.9.2 \citep{vandervelden20},
Emcee v3.1.6 \citep{foremanmackey13},
iSpec v2023.08.04 \citep{blancocuaresma14,blancucuaresma19}
Jax v0.4.32 \citep{jax2018},
Matplotlib v3.9.2 \citep{Hunter:2007}, 
MESA r23.05.1 \citep{paxton11,paxton13,paxton15,paxton18,paxton19,jermyn2023}, 
NumPy v1.26.4 \citep{numpy:2011,harris20}, 
NumPyro v0.15.2 \citep{phan19,bingham19}
Pandas v2.2.2 \citep{pandas:2010}, 
PySynphot v1.0.0 \citep{pysynphot13},
Python 3.7.8, 
Scikit-learn v1.6.1 \citep{scikit-learn11},
Scipy v1.14.1 \citep{scipy:2001,scipy:2020}
}

\newpage

\bibliography{bib}
\bibliographystyle{aasjournal}

\appendix
\restartappendixnumbering

\section{Updated Stellar Parameters}\label{app:updatedparams}

As mentioned in \S{\ref{sec:sample}}, we updated our template fitting procedure from that used by \citetalias{chen24} in order to incorporate our measurements of the additional turbulent line broadening. To do this in sampling, after loading the ATLAS9 model corresponding to the given effective temperature and surface gravity but before scaling to the radius and applying extinction, we broaden the model flux first with a Gaussian with a width equal to the measured $\vmicro$ value, then with a macroturbulence kernel with the measured $\vturb$. This is accomplished by following the numerical recipes adopted in the {\sc TSFitPy} package \citep{gerber23,storm23}.

We also made minor tweaks to the sampling procedure. The first was an update to the likelihood function. The function used by \citetalias{chen24} was defined as
\begin{equation}
    \ln \mathcal{L} = -\frac{1}{2}\sum_i \frac{(f_i-\hat{f}_i)^2}{\sigma_i^2}
\end{equation}
where $f_i$ and $\sigma_i$ are the flux and uncertainty in the $i$th wavelength bin, and $\hat{f}_i$ is the model flux associated with a given set of parameters. In order to account for the possibility that the errors on the flux measurements of our spectra are underestimated, we introduce a new parameter, $s$, which corresponds to an additional fractional uncertainty relative to the model. We now define
\begin{equation}
    \sigma_{i,{\rm new}}^2 = \sigma_i^2 + s^2\hat{f}_i^2
\end{equation}
and update the likelihood function to
\begin{equation}
    \ln \mathcal{L} = -\frac{1}{2}\sum_i \frac{(f_i-\hat{f}_i)^2}{\sigma_{i,{\rm new}}^2} + \ln (2\pi\sigma_{i,{\rm new}}^2)
\end{equation}
where the second term is a regularization term; without it, the likelihood function would be maximized for $s\rightarrow\infty$.

In practice, we perform the sampling on $s$ in log-space, and set a uniform prior $\log s \sim \mathcal{U}(-10, 0)$. For completeness, we place uniform priors $T_{\rm eff}/\rm{K} \sim \mathcal{U}(3500 ,10500)$, $\log g \sim \mathcal{U}(-1, 4)$, $E(B-V)/\rm{mag}\sim \mathcal{U}(0, 2.5)$, $R/R_\odot\sim\mathcal{U}(0.5,1000)$, and $v_{r}/{\rm km\ s}^{-1}\sim\mathcal{N}(v_{r,\rm{lit}}, 30)$ where $v_{r,\rm{lit}}$ is the literature radial velocity. 

The second tweak was to identify erroneous points caused by our order stitching procedure; at low signal to noise (i.e., in the blue), mismatches between orders would cause individual pixels in the overlap region to have extreme flux values. To identify these pixels, we numerically computed the gradient of the spectrum, and removed all points for which the absolute value of the gradient exceeded $10^{-15}$ erg s$^{-1}$ cm$^{-2}$ \AA$^{-1}$ pixel$^{-1}$. This value was identified by manually inspecting the distribution of gradient values, and selecting a cutoff that only removed bad pixels, without excluding pixels with large gradients due to the typical noise in the data.

Finally, in order to ensure convergence of our sampler, we adopt a similar auto-correlation approach as described in \S\ref{subsec:profilefit} and \S\ref{subsec:slfv}. Though exhaustive, this approach did significantly increase the sampling time. To account for this, we changed the sampling approach such that the sampler would randomly propose a step using a differential-evolution procedure from \citet{nelson14} for 80\% of steps, and using a snooker proposal from \citet{terbraak08} 20\% of the time. Otherwise, our procedure is nearly identical to that described in \citetalias{chen24}.

\end{document}